\documentclass[11pt,prd,nofootinbib,reprint,superscriptaddress,longbibliography]{revtex4-2}
\usepackage{amsmath, amssymb, amsthm, graphicx, epsfig, fancyhdr,epsfig,multirow}
\usepackage[utf8]{inputenc}
\usepackage{amsmath}
\usepackage{amsfonts}
\usepackage{amssymb}
\usepackage{xcolor}
\usepackage{comment}
\usepackage{subcaption}
\usepackage[normalem]{ulem}
\usepackage{tabularx}
\usepackage{comment}
\usepackage[left=2cm,right=2cm,top=2cm,bottom=2cm]{geometry}
\usepackage[font=small,labelfont=bf,textfont=it]{caption}

\DeclareUnicodeCharacter{2009}{\,} 
\DeclareUnicodeCharacter{2212}{-}
\newcommand{\nn}{\nonumber}

\usepackage{color}
 
\begin{document}
\title{Blazar boosted Dark Matter - direct detection constraints on $\sigma_{e\chi}$ :\\ Role of energy dependent cross sections}
\author{Supritha Bhowmick}
\email{supritha.bhowmick@students.iiserpune.ac.in}
\affiliation{Department of Physics, Indian Institute of Science Education and Research Pune, India}
\author{Diptimoy Ghosh}  
\email{diptimoy.ghosh@iiserpune.ac.in}
\affiliation{Department of Physics, Indian Institute of Science Education and Research Pune, India}
\author{Divya Sachdeva}
\email{divya.sachdeva@phys.ens.fr}
\affiliation{Laboratoire de Physique de l’Ecole Normale Sup\'erieure, CNRS, Universit\'e
PSL, Sorbonne Universit\'es, 24 rue Lhomond, 75005 Paris, France
}

\begin{abstract}
Elastic collisions with relativistic electrons from the blazar's jet can accelerate dark matter (DM) particles in the DM spike surrounding the supermassive black hole at its center. This can allow one to set stringent limits on the DM-electron scattering cross section ($\bar{\sigma}_{e\chi}$) for DM masses less than 100 MeV.  We consider DM particles boosted by energetic electrons in the jets of the blazars TXS 0506+056 and BL Lacertae. Both vector and scalar mediators for the scattering of electron and electrophilic fermionic DM are studied. We highlight that the ensuing energy dependence of the S-matrix for the corresponding Lorentz structure of the vertex significantly modifies the constraints. We find that the revised exclusion limits are orders of magnitude stronger than the equivalent results for the simple constant cross section assumption. Our limits are also assessed for the less cuspy spike. 
\end{abstract}

\maketitle
\section{Introduction}
The Cold Dark Matter (CDM) provides a compelling explanation for a broad range of observations, including rotation curves in spiral galaxies, gravitational micro-lensing, cluster collisions (the Bullet Cluster), and temperature anisotropy in the spectrum of cosmic microwave background radiation. To that end, a variety of particle physics models predict a feeble interaction between SM and DM, which can be investigated using Direct detection (DD) experiments. The DD experiments identify the nuclear or electronic recoils produced by the scattering between DM and the detector’s (target) nuclei or electrons. The average velocity of DM particles in the solar vicinity, however, restricts the amount of energy that may be deposited in a detector. For example: detectors like {\sc{ Xenon1T}} can detect DM mass $m_\chi\sim\mathcal{O}(1~{\rm MeV}$), corresponding to electronic recoil of $\sim \mathcal{O}(1~{\rm keV})$~\cite{XENON:2022ltv}. The neutrino detectors like {\sc{Super-K}} are sensitive to recoil energy threshold of $\sim \mathcal{O}(1~{\text{MeV}})$~\cite{Super-Kamiokande:2011lwo,Super-Kamiokande:2017dch} leading to the smallest accessible DM mass of $\mathcal{O}(1~{\text{GeV}})$ \footnote{Fermionic DM absorption models~\cite{Dror:2020czw,Dror:2019onn,Dror:2019dib} allow {\sc{ Xenon1T}} and {\sc{Super-K}} to probe masses down to $\sim \mathcal{O}(10~{\rm keV})$ and $\sim \mathcal{O}(1~{\rm MeV})$ respectively}. Thus, these detectors appear to have a limited range for detecting lighter DM particles. Since these observations have been unfavourable towards confirmed detection of DM, it is critical to develop methods for probing the sub-GeV/MeV mass range.

The reach of these experiments has been extended to DM masses well below 1 GeV in recent years, thanks to the novel idea of boosting the halo DM through its interaction with the SM particles via cosmic rays~ \cite{Bringmann:2018cvk,Cappiello:2018hsu,Cappiello:2019qsw,Bringmann:2018cvk,Ema:2018bih,Cappiello:2018hsu,Cappiello:2019qsw,Dent:2020syp,Jho:2020sku,Bramante:2021dyx,Farzan:2014gza,Arguelles:2017atb,Yin:2018yjn,Jho:2021rmn,Das:2021lcr,PhysRevD.105.103029,Cao:2020bwd,Dent:2020syp,Ema:2020ulo,Xia:2022tid,Elor:2021swj,Maity:2022exk}, primordial black holes~\cite{Calabrese:2022rfa,Calabrese:2021src}, diffuse Supernova Neutrino Background (DSNB)~\cite{Farzan:2014gza,Arguelles:2017atb,Yin:2018yjn,Jho:2021rmn,Das:2021lcr}, and blazars~\cite{Wang:2021jic,Granelli:2022ysi}.Even though the boosted DM flux is substantial and DM particles are (semi)relativistic, the population of the up-scattered subcomponent of DM is significantly lower than the galactic DM population. Hence sensitivity for boosted DM is achieved at cross sections much larger than unboosted scenario.

In this paper, we consider the blazar boosted DM, which was proposed in Ref.~\cite{Wang:2021jic,Granelli:2022ysi} in the context of fermionic DM. The presence of a supermassive Black Hole (BH) at the blazar center, which provides a dense DM population compensates for the blazar's large distance from Earth by producing DM flux that is stronger than that from galactic CRs. The existing literature, however, assumes that DM interaction cross-sections are independent of DM energy. Although this simple assumption makes calculations easier, it's not physically realistic. It would also be an incorrect approximation to make, notably in scenarios where a significant DM flux becomes relativistic after being scattered by energetic particles. Some of the previously mentioned works~\cite{Dent:2019krz,Cao:2020bwd,Dent:2020syp,Ema:2020ulo,Xia:2022tid} for cosmic ray boosted DM, have already discovered that the limits for energy-dependent cross-section differ by orders of magnitude from those obtained under the assumption of a constant cross section. 

The notion of an energy-dependent scattering cross-section is thus primarily investigated in the present work by taking into account electrophilic fermionic DM that has been boosted by energetic electrons from blazars. To constrain the scattering cross-section, we use electron recoil measurements in Super-Kamiokande. This work is organized as follows. We discuss the spectrum of energetic particles in the blazar jets in Section II, and describe DM density profiles in Section III. In Section IV, we estimate the Blazar boosted DM (BBDM) flux and compute the event rate. In Section V, we present simplified DM models. We present the main results of our paper in Section VI, i.e., the energy dependent exclusion bound from BBDM-electron scattering, and in Section VII, we summarize and conclude.

\section{Blazar Jet Spectrum} 
Blazars are characterized by a non-thermal spectral energy distribution (SED). This spectrum has a low energy peak in the infra-red or X-ray region, which has been accepted to be due to synchrotron emission of electrons in the jet \cite{Abdo:2009iq}. Another peak at $\gamma$-ray frequencies could be due to highly relativistic protons~\cite{Keivani:2018rnh,Cerruti:2018tmc,Rodrigues:2018tku,Xue:2019txw,Petropoulou:2019zqp}, as motivated by the recent IceCube detection~\cite{IceCube:2018dnn,IceCube:2018cha,Padovani:2018acg} of a high energy neutrino from TXS 0506+056 blazar. Since DM considered in this work is electrophilic, at tree level it can only interact with electrons. Therefore, we are only concerned with the blazar jets' electron spectrum. 

We follow the procedure laid out in Ref.~\cite{Wang:2021jic,Granelli:2022ysi} to compute the spectrum of the energetic electrons in the blazar jets, assuming~``Blob geometry" model~\cite{book1}. In this model, the energetic particles in the blazar jets move isotropically in a ``blob" frame, as the blob traverses outwards along the jet axis. The Lorentz boost factor of the blob is given by $\Gamma_B = (1-\beta_B^2)^{-1/2}$, where $\beta_B$ is the blob's propagation speed. The inclination of the jet axis with respect to the line of sight (LOS) is taken to be $\theta_{\text{LOS}}$.

In the blob frame, the energetic electrons follow a power law distribution with a high and a low energy cutoff ($\gamma '_{\text{max},e}$ and $\gamma '_{\text{min},e}$ respectively). This spectrum can then be frame transformed to the observer's rest frame (for details of the derivation, see \cite{Wang:2021jic}), given by :
\begin{eqnarray}
\frac{d\Gamma_e}{dT_e d\Omega}=\frac{c_e }{4\pi} \Gamma_B^{-\alpha_e} \left( 1+ \frac{T_e}{m_e} \right)^{-\alpha_e} \nonumber \\
\times \frac{\beta_e (1-\beta_e \beta_B \mu)^{-\alpha_e}}{\sqrt{\left(1-\beta_e \beta_B \mu\right)^2 -\left(1-\beta_e^2\right)\left(1-\beta_B^2\right)}} \label{eq:blazar_jet_spectrum}
\end{eqnarray}

where $m_e$ and $T_e$ are the mass and kinetic energy of the electron respectively. The speed of the electrons is given by $\beta_e = \left(1- m_e^2/(T_e + m_e)^2 \right)^{1/2}$. The doppler factor for the blob frame is $\mathcal{D}=\left( \Gamma_B(1-\beta_B \cos \theta_{\text{LOS}}) \right)^{-1}$. $\alpha_e$ is the power index of the electron spectrum in the blob frame. $\mu$ is the cosine of the angle between direction of the electron's motion and the jet axis. It is related to the scattering angle in the blob frame ($\bar{\mu}_s$) by \cite{Bringmann:2018cvk,Wang:2021jic} :
\begin{eqnarray}
\mu(\bar{\mu}_s,\phi_s)=\bar{\mu}_s \cos \theta_{\text{LOS}} + \sin \phi_s \sin \theta_{\text{LOS}} \sqrt{1-\bar{\mu}_s^2}
\label{eq:mu-mus}
\end{eqnarray}

where $\phi_s$ is the azimuth with respect to the LOS. $\bar{\mu}_s$ is related to the kinetic energy of the blazar jet electron and the kinetic energy ($T_{\chi}$) transferred to the DM, as follows :
\begin{eqnarray}
\bar{\mu}_s (T_e,T_{\chi})=\left[ 1+\frac{T_{\chi}^{\text{max}} -T_{\chi}}{T_{\chi} }\frac{(m_e +m_{\chi})^2 + 2 m_{\chi} T_e}{(T_e + m_e + m_{\chi})^2 } \right]^{-1/2}
\end{eqnarray} 

Now, $c_e$ is a normalisation constant which is determined from the blazar jet electron luminosity ($L_e$), where the latter depends on $c_e$ as~\cite{Gorchtein:2010xa,Wang:2021jic} :
\begin{eqnarray}
L_e = c_e m_e^2 \Gamma_B^2 \int _{\gamma '_{\text{min},e}}^{\gamma '_{\text{max},e}} \left(\gamma'_e \right)^{1-\alpha_e} d\gamma_e' ~,
\end{eqnarray} 

and thus $c_e$ is simply given by :
\begin{widetext}
\begin{align}
c_e = \frac{L_e}{m_e^2 \Gamma_B^2} \times 
\begin{cases}
     \mbox{ $\left( 2-\alpha_e \right) /\left[ \left(\gamma'_{\text{max},e}\right)^{2-\alpha_e}
-\left(\gamma'_{\text{min},e}\right)^{2-\alpha_e}
\right] $ } & \text{if 
$\alpha_e \neq 2$ ;} \\
   \mbox{\tiny\( \)} & \mbox{\tiny\( \)} \\
      1/\log \left(\gamma'_{\text{max},e} / \gamma'_{\text{min},e} \right)  & \text{if $\alpha_e =2$. }
    \end{cases}
\end{align}
\end{widetext}

\begin{figure}[!h]
\hspace*{-6 mm}
\includegraphics[scale=0.6]{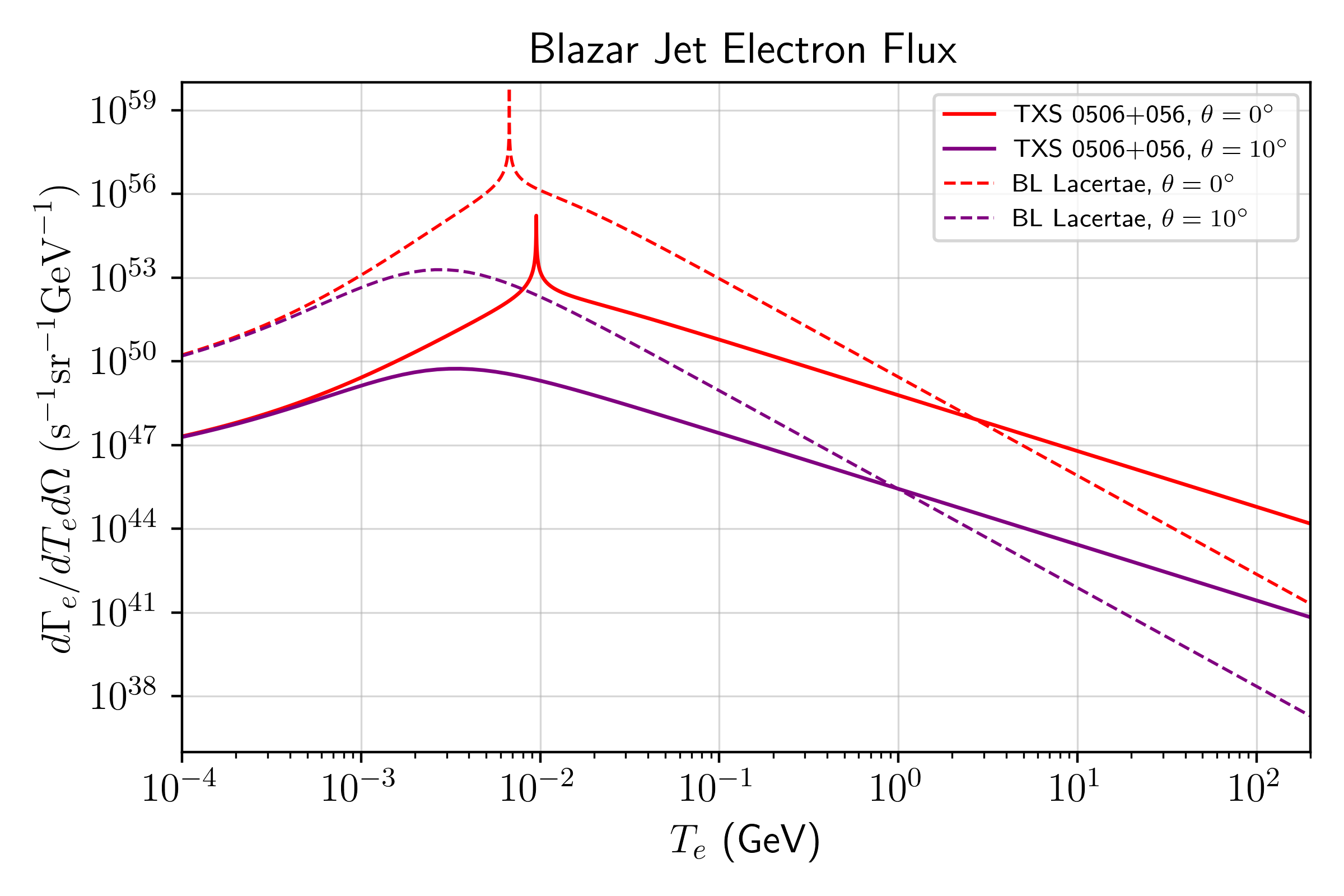}
\caption{The electron spectrum in the observer's frame is plotted above, for the blazars TXS 0506+056 (solid lines) and BL Lacertae (dashed lines). The spectrum is shown for two different polar angles : $\theta = 0^{\circ}$ (in red), $\theta=10^{\circ}$ (purple). For larger kinetic energies ($T_e \gtrsim 10~\rm{GeV}$), the electron flux from TXS 0506+056 blazar exceeds the flux from BL Lacertae.}
\label{fig:Blazar_jet_flux}
\end{figure}

The parameters $\gamma'_{\text{min},e}$, $\gamma'_{\text{max},e}$, $\alpha_e$, $L_e$ and $\mathcal{D}$ are fitted to the SED of a blazar. The doppler factor is assumed to be either $2\Gamma_B$ or $\Gamma_B$. These two cases correspond to TXS 0506+056 ($\theta_{\text{LOS}}=0$) and BL Lacertae ($\theta_{\text{LOS}}\sim 3.82^\circ$). All the parameters required to find the blazar jet spectrum of TXS 0506+056 and BL Lacertae, along with the blazar redshift and luminosity distance ($d_L$) are mentioned in Table \ref{table:model_param}. The electron spectrum is plotted in Fig.~\ref{fig:Blazar_jet_flux}.
\begin{table}[htb]
\resizebox {\columnwidth}{!}{
\begin{tabularx} {0.4\textwidth}{ 
   >{\centering\arraybackslash}X 
   >{\centering\arraybackslash}X 
   >{\centering\arraybackslash}X }
\hline 
Parameter & TXS 0506+056 & BL Lacertae \\
\hline \hline
Redshift & 0.337 & 0.069 \\
$d_L$  & $1835.4~\text{Mpc}$ & $322.7~\text{Mpc}$ \\
$M_{\text{BH}}$ & $3.09 \times 10^8~M_{\odot}$ & $8.65 \times 10^7~M_{\odot}$ \\
$\Gamma_B$ & $20$ & $15$ \\
$\theta_{\text{LOS}}$ & $0^{\circ}$ & $3.82^{\circ}$ \\
$\alpha_e$ & $2$ & $3.5$ \\
$\left(\gamma'_{\text{min},e},\gamma'_{\text{max},e}\right)$ & $(500,1.3 \times 10^4)$
& $(700,1.5\times10^4)$ \\
$L_e~(\text{erg}/\text{s})$ & $ 1.32\times10^{44}$ & $8.7\times10^{42}$ \\ 
\end{tabularx}
}
\caption{Model parameters for TXS 0506+056 \cite{Cerruti:2018tmc} and BL Lacertae blazars \cite{Boettcher:2013wxa}. } \label{table:model_param}
\end{table}

\section{DM density profile}
N-body simulations and observations are not sensitive at subparsec sizes, thus the DM distribution near Galactic center is not well known. The central supermassive black hole (SMBH) can have a considerable impact on DM density if the SMBH grows adiabatically, i.e., on a  timescale much longer than its dynamical timescale. The DM density in a region corresponding to the sphere of gravitational influence of the black hole (BH) is expected to be significantly enhanced~\cite{Gondolo:1999ef}. This results in a morphological feature known as a DM spike, which corresponds to a DM profile with a power law scaling $\rho(r) \propto r^{-\gamma_{\textrm{sp}}}$~\cite{Gondolo:1999ef}. Here $\gamma_\textrm{sp}=\frac{9-2\gamma}{4-\gamma}$ commonly ranges from $2.25$ to $2.5$, depending on the slope of the initial DM halo distribution, $\gamma$. In this work, we assume the initial central DM profile is Navarro-Frenk-White, $\gamma\,=\,1$. Also, for DM annihilating with cross-section $\langle\sigma v\rangle_{\text{ann.}}$, the innermost region of the DM spike is depleted because DM particles annihilate efficiently on account of high DM density, leading to an “annihilation plateau” density given by
\begin{equation}
    \rho_{\rm sat} = \frac{m_\chi}{\langle\sigma v\rangle_{\text{ann.}} t_{\rm BH}},
\end{equation}
where $t_\textrm{BH}\sim\,10^9\,{\rm yrs}$ \cite{Granelli:2022ysi} is the age of the BH. The DM density profile in such a spike is given by
\begin{align}
\label{eq:spikedensity}
    \rho(r)= \left\{
        \begin{array}{ll}
            0 & \quad r<4 R_S \\
            \rho_\textrm{sat} & \quad 4 R_S \leq r < R_\textrm{sat} \\
            \mathcal{N}_1r^{-\gamma_\textrm{sp}} & \quad R_\textrm{sat} \leq r < R_\textrm{sp} \\
            \mathcal{N}_2 r^{-\gamma}& \quad r \geq R_\textrm{sp}
        \end{array} ,
    \right.
\end{align}
where $R_S=2GM/c^2$ is the Schwarzchild radius of the BH, $R_{\rm sp}\,=\,10^5\,R_s$ is the radius of the spike~\cite{Gorchtein:2010xa} and {$\rho (r)$ goes to zero in the region $r<4 R_s$ due to DM particles being captured by the SMBH. The saturation density $\rho_{\rm sat}$ and the saturation radius $R_{\rm sat}$
are related by the equality $\rho(R_{\rm sat})=\rho_{\rm sat}$. In this work, the normalization $\mathcal{N}_1$ of $\rho$ is determined by observing that the mass of the spike is of the same order as $M_{\rm BH}$ within the spike radius~\cite{Ullio:2001fb} and $\mathcal{N}_2$ is determined by observing the continuity of the profile at $R_{\rm sp}$. However, this choice of normalisation ($\mathcal{N}_1$) is only an optimistic upper limit, and there could be way less DM close to the BH. We show, in Appendix :\ref{appendix:A}, how the main results of our work (i.e. the exclusion limits in Figs.~\ref{fig:ExclBound_BMP1},\ref{fig:ExclBound_BMP2} ) scale for the choice of different normalisations $\mathcal{N}_1$.

For a pre-existent DM halo with $\gamma=1$, the final DM profile near BH corresponds to $\gamma_{\text{sp}} = 7/3$. A more realistic model was obtained in Ref.~\cite{Bertone:2005hw}, where the time-evolution of dark matter distribution was investigated on sub-parsec scales. This implied softening of the DM density spike, due to scattering of DM by stars and capture of DM particles by the SMBH, dampening it to $\gamma_{\text{sp}} = 3/2$. Thus, in this work, we consider both the DM profile parameters $\gamma_{\text{sp}} = 7/3$ and $\gamma_{\text{sp}} = 3/2$ along with two extreme values of $\langle\sigma v\rangle_{\text{ann.}}$. We define these as
\begin{itemize}
\item[]Profile 1: $\rho(R_\textrm{sat} \leq r < R_\textrm{sp})\,=\,\mathcal{N}_1r^{-7/3}$
\item[]Profile 2: $\rho(R_\textrm{sat} \leq r < R_\textrm{sp})\,=\,\mathcal{N}_1r^{-3/2}$.
\end{itemize}

It should be noted that the formation of DM spike can be influenced by various factors, including mergers with other galaxies and the distance between the BH and the Galactic center~\cite{Ullio:2001fb,Bertone:2005hw}. While some of these uncertainties have been addressed in studies of the formation of the DM spike in the Milky Way, no such studies exist for the galaxy considered in our paper. Therefore, we provide a preliminary analysis of the potential effects of DM close to a BH in a galactic center.

For each of these profiles, the two benchmark points (BMPs) are defined as:
\begin{itemize}
\item[]BMP 1: No DM annihilation, i.e., $\langle\sigma v\rangle_{\text{ann.}}|_{\rm tot} = 0$.  Here, we assume that the DM annihilation is forbidden by some symmetry.

\item[]BMP 2:  $\langle\sigma v\rangle_{\text{ann.}}|_{\rm tot} = 3 \times 10^{-26} ~\text{cm}^3 \text{s}^{-1}$, thermal relic cross-section.
\end{itemize}
Another quantity relevant to the computation of the BBDM flux is the line of sight (LOS) integral of DM density around the blazar. This provides a measure of the number of DM particles being boosted by the blazar. At a certain distance $r$ from the blazar, it is defined as}
\begin{eqnarray}
\Sigma_{\text{LOS}}(r)=\int_{r_{\text{min}}} ^{r} \rho (r') dr'
\end{eqnarray}
where $r_{\text{min}}$ is the distance from the SMBH from where the blazar jet starts. To get a measure of all boosted DM particles, we want the LOS integral at large distances ($r>>10^5 R_S$), and we define $\Sigma_{\text{LOS}} ^{\text{tot}} = \Sigma_{\text{LOS}}(r>>10^5 R_S)$.

In this work, we will study BBDM flux from TXS 0506+056 and BL Lacertae, and for these blazars, $r_{\text{min}}$ lies within $100 R_S$ \cite{Boettcher:2013wxa,Cerruti:2018tmc,Gorchtein:2010xa}. We take $r_{\text{min}}=4 R_S$, noting that $\Sigma_{\text{LOS}} ^{\text{tot}}$ is independent of choice of $r_{\text{min}}$ for models which allow for DM pair annihilation. For no DM annihilation, $\Sigma_{\text{LOS}} ^{\text{tot}}$ decreases by an order of magnitude when $r_{\text{min}}$ is changed to $100 R_S$.

\begin{figure}[!h]
\begin{subfigure}{\linewidth}
\hspace*{-6 mm}
\includegraphics[scale=0.6]{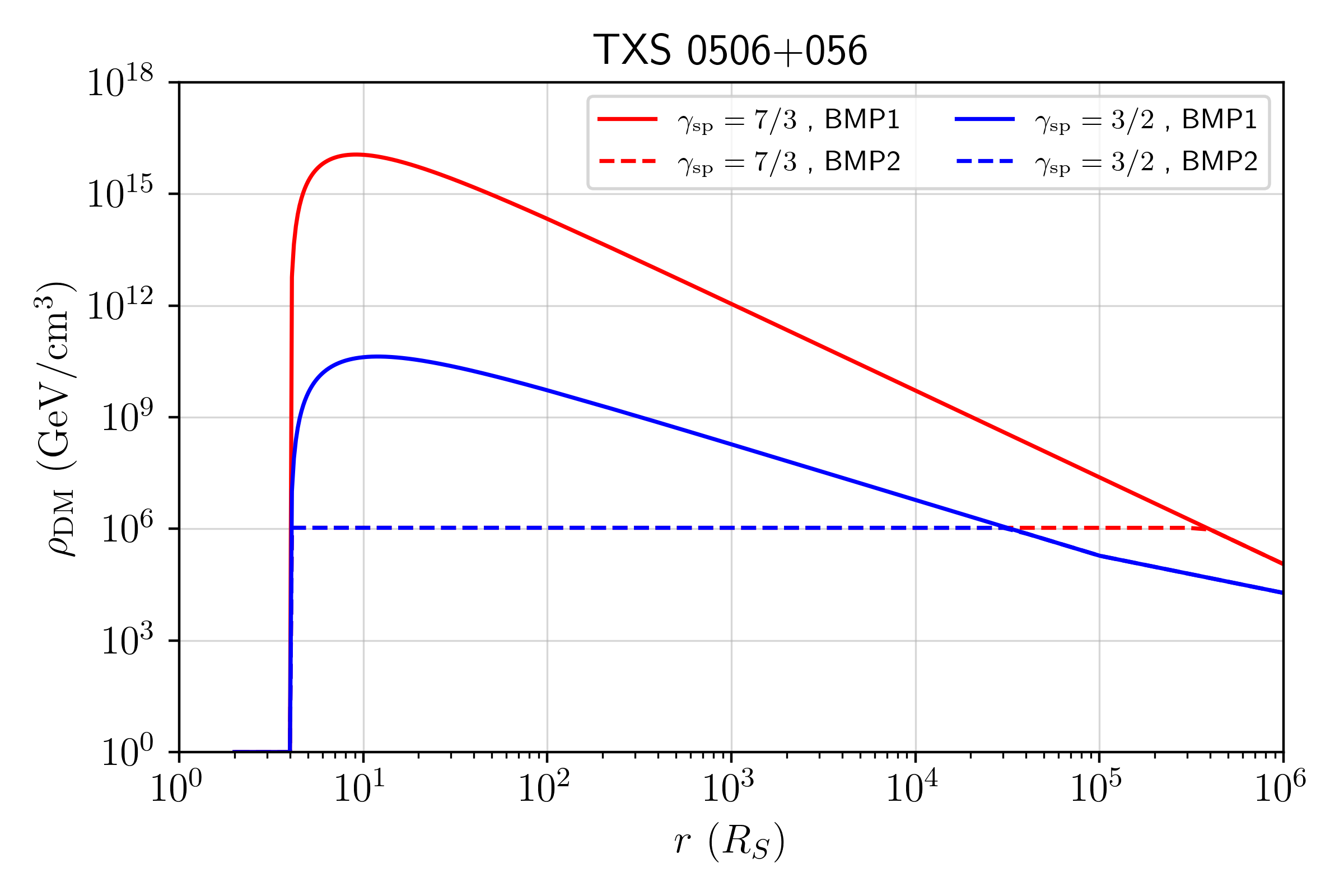}
\caption{DM Density Profile}
\label{fig:densityprof}
\end{subfigure}
\begin{subfigure}{\linewidth}
\hspace*{-6 mm}
\includegraphics[scale=0.6]{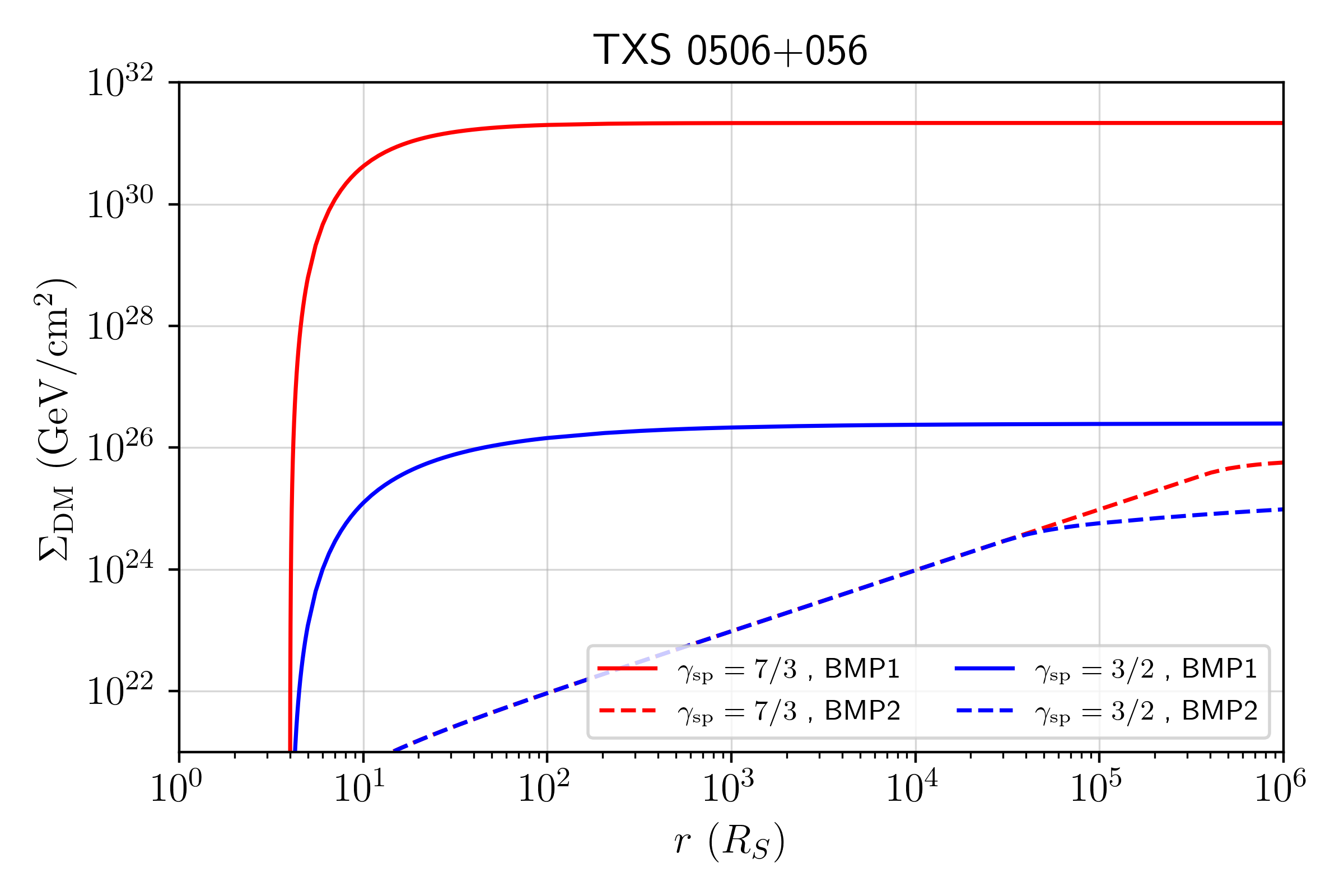}
\caption{LOS integral profile}
\label{fig:LOSprof}
\end{subfigure}
\caption{The profiles of $\rho_{\rm DM}$ (Fig.~\ref{fig:densityprof}) and $\Sigma_{\rm DM}$ (Fig.~\ref{fig:LOSprof}) are plotted above for TXS 0506$+$056 blazar parameters. The DM mass chosen for these figures is $m_\chi = 1~{\rm MeV}$. Profile 1 (red) and Profile 2 (blue) are plotted for BMP 1 (solid curve) and BMP 2 (dashed curve). BL Lacertae, on the other hand, is less massive than TXS 0506+056, and yields a larger spike (for BMP1) at a smaller distance from the BH (See text for more details).}
\label{fig:DMDensity}
\end{figure}

The DM density and L.O.S. integral profiles are plotted in Fig~\ref{fig:DMDensity} for two benchmark points. For both DM density profiles, BMP1 yields a larger spike and a larger LOS integral ($\Sigma_{\text{LOS}} ^{\text{tot}}$). This results in a larger BBDM flux and consequently a stronger exclusion bound on DM-electron interaction cross section. Hence, one can expect models with no DM annihilation to yield better bounds. Moreover, even though the LOS integral for the idealistic spike (Profile 1) is very large and thus results in substantial BBDM flux, the more realistic profile (Profile 2) with a softer spike would lead to a much smaller LOS integral, and hence a much weaker exclusion bound.

\section{Blazar Boosted Dark Matter Flux and Event Rate} 
DM particles are boosted via elastic collisions with the relativistic electrons in the blazar jet. The DM differential flux resulting out of collision with the electrons is obtained as follows :
\begin{eqnarray}
\frac{d\phi_{\chi}}{dT_{\chi}} = \frac{\Sigma_{\text{DM}}^{\text{tot}}}{2\pi m_{\chi}d_L^2} \int_0^{2\pi}d\phi_s \int_{T_e^{\text{min}}(T_{\chi},\phi_s)}^{T_e^{\text{max}}(T_{\chi},\phi_s)} dT_e \nonumber \\
\times \frac{d\sigma_{\chi e}}{dT_{\chi}}  \frac{d\Gamma_e}{dT_e d\Omega}~,
\label{eqn:BBDM flux}
\end{eqnarray} 
where $\sigma_{\chi e}$ is the DM-electron interaction cross section. The integration over $\phi_s$ becomes trivial in case of TXS 0506+056, where the system is symmetric about LOS, and we can simply set $\mu=\bar{\mu}_s$ (from Eqn.~\eqref{eq:mu-mus}).  

The maximal kinetic energy of the blazar jet electrons along LOS is given by $T_{e,\text{jet}}^{\text{max}}=m_e \left( \gamma'_{\text{max},e} ~\Gamma_B^{-1} (1-\beta_B \cos \theta_{\text{LOS}})^{-1} -1 \right)$. This is set as the upper bound of the integral on $T_e$ in Eqn.~\eqref{eqn:BBDM flux}. The lower bound is set by the minimum kinetic energy required for scattering, given by 
\begin{eqnarray}
T_e^{\text{min}}=\left(\frac{T_{\chi}}{2} - m_e \right) \left[1 \pm \sqrt{1+ 
\frac{2 T_{\chi} (m_e +m_{\chi})^2}{m_{\chi} (T_{\chi}-2m_e)^2}}~\right]~,
\label{eq:Temin}
\end{eqnarray}
with $+$ and $-$ applicable for $T_{\chi}>2m_e$ and $T_{\chi}<2m_e$ respectively. However, the kinetic energy of the slowest electrons in the blazar jets could be larger than $T_e^{\text{min}}$. In such a case, the kinetic energy of the least energetic electron in the jet, given by $T_{e,\text{jet}}^{\text{min}} = m_e \left( \gamma'_{\text{min},e} ~\Gamma_B^{-1} (1-\beta_B \cos \theta_{\text{LOS}})^{-1} -1 \right)$, sets the lower bound of the integral in Eqn.~\eqref{eqn:BBDM flux}. 

The differential cross section ( $d\sigma_{\chi e}/dT_{\chi}$ ) of the DM-blazar jet electron interaction is given by,
\begin{equation}
\frac{d \sigma_{\chi e}}{d T_{\chi}}=\frac{|\mathcal{M}|^{2}}{16 \pi s_{e}} \frac{1}{T_{\chi}^{\max }}
\label{eqn:diffxsec}
\end{equation}
where $\mathcal{M}$ is the interaction matrix element, a function of $T_{\chi}$ and $T_{e}$. $s_{e}$ is the centre of momentum energy for the electron-DM collision given by :
\begin{eqnarray}
s_{e}=\left(m_\chi +m_e\right)^2 + 2m_\chi T_{\text{e}}~,
\label{eqn:se}
\end{eqnarray} 
and $T_{\chi}^{\max }$ is the maximum kinetic energy that can be imparted to a DM particle by a blazar jet electron of energy $T_e$ is given by :
\begin{eqnarray}
T^{\rm max}_\chi  &=& \frac{T_{e}^{2}+2 m_{e} T_{e}}{T_{e}+\left(m_{e}+m_{\chi}\right)^{2} /\left(2 m_{\chi}\right)}
\label{eqn:Tchimax} 
\end{eqnarray}

TXS 0506+056 blazar is more massive as compared to BL Lacertae. Keeping this in mind allows us to qualitatively compare the DM density spikes and LOS integrals from these two blazars. For BMP1, the normalisation factor $\mathcal{N}_1$ of the density profile is proportional to $M_{\text{BH}}^{1/3}$ for Profile 1, while for Profile 2, $\mathcal{N}_1$ is proportional to $M_{\text{BH}}^{-1/2}$. The LOS values for both profiles vary as $M_{\text{BH}}^{-1}$, hence BL Lac yields a larger spike as well as a larger LOS value for BMP1. However, for BMP2, most of the DM density profile is determined by $\rho_{\text{sat}}$, which is independent of $M_{\text{BH}}$. The LOS values of the two blazars are thus not significantly different for BMP2. TXS 0506+056 is also further away from us as compared to BL Lacertae. Overall, the contribution to the DM flux coming from the factors ($\Sigma_{\text{LOS}}/ d_L^2 $) outside the integrals in Eqn.~\eqref{eqn:BBDM flux} is expected to be significantly larger for BL Lacertae than TXS 0506+056.

\begin{figure}[!h]
\begin{subfigure}{\linewidth}
\hspace*{-6 mm}
\includegraphics[scale=0.6]{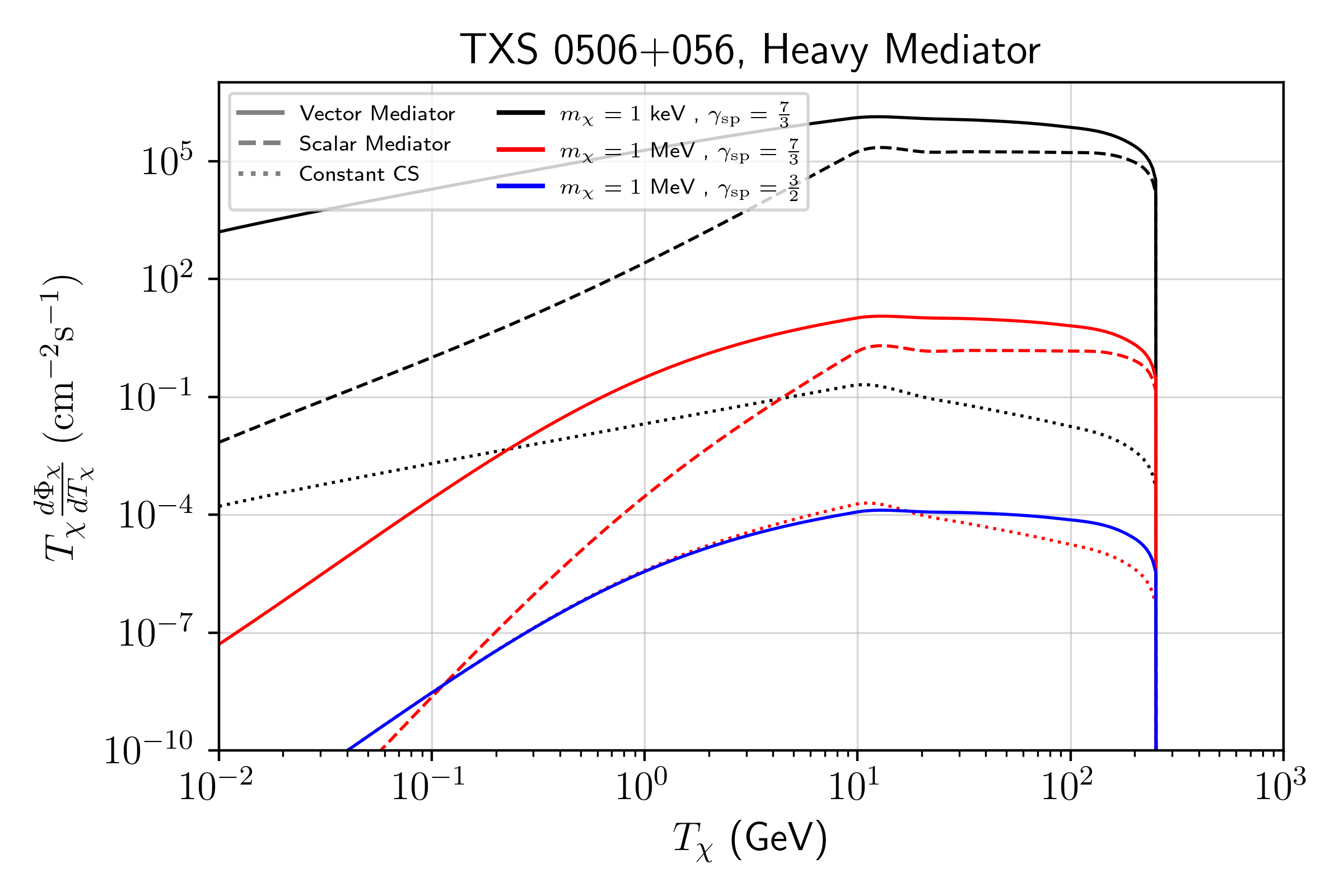}
\caption{BBDM flux for heavy mediator scenario}
\label{fig:heavymed_TXS}
\end{subfigure}
\begin{subfigure}{\linewidth}
\hspace*{-6 mm}
\includegraphics[scale=0.6]{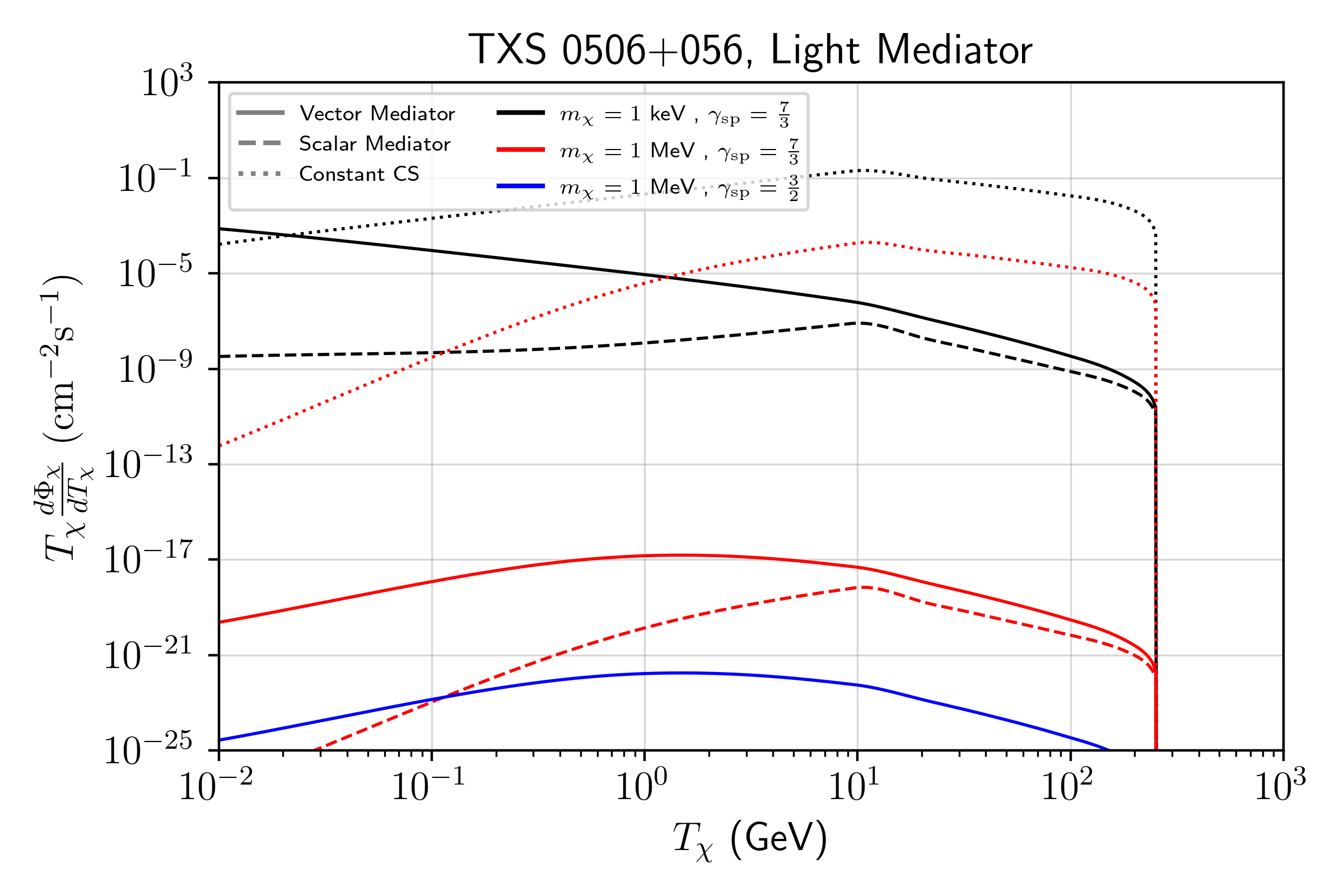}
\caption{BBDM flux for light mediator scenario}
\label{fig:lightmed_TXS}
\end{subfigure}
\caption{Flux of DM particles, boosted by energetic electrons in the jets of TXS 0506+056 blazar, is plotted above, for heavy (\ref{fig:heavymed_TXS}) and light (\ref{fig:lightmed_TXS}) mediators.  The parameters chosen for the above plots are $\bar{\sigma}_{e\chi}=10^{-30}~\text{cm}^2$ and BMP1. The vector and the scalar mediator cases have been plotted in solid and dashed lines respectively. For comparision, DM flux for constant cross section scenario have also been plotted in dotted lines. Two DM masses have been considered, $m_\chi =1~\text{keV}$ (plotted in black) and $m_\chi = 1~\text{MeV}$ (plotted in red) for DM density Profile 1 (i.e. $\gamma_{\text{sp}} = 7/3$). To avoid overcrowding, only vector mediator case is considered for Profile 2 (i.e. $\gamma_{\text{sp}} = 3/2$), and BBDM flux is plotted (in blue) corresponding to DM mass $m_\chi = 1~\text{MeV}$. Clearly, Profile 2 yields a smaller BBDM flux as compared to Profile 1.}
\label{fig:BBDM Flux_TXS}
\end{figure}

\begin{figure}[!h]
\begin{subfigure}{\linewidth}
\hspace*{-6 mm}
\includegraphics[scale=0.6]{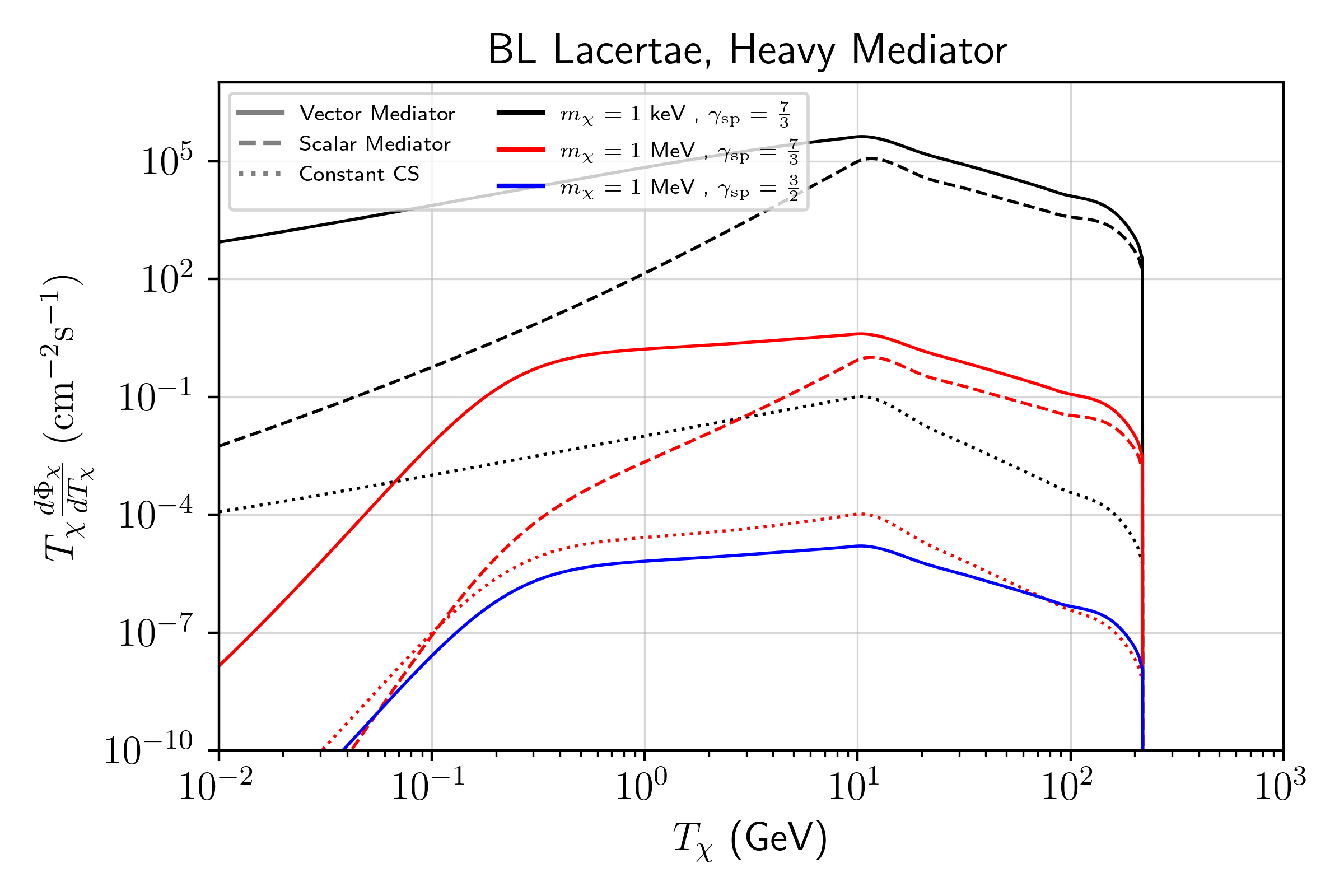}
\caption{BBDM flux for heavy mediator scenario}
\label{fig:heavymed_BLLac}
\end{subfigure}
\begin{subfigure}{\linewidth}
\hspace*{-6 mm}
\includegraphics[scale=0.6]{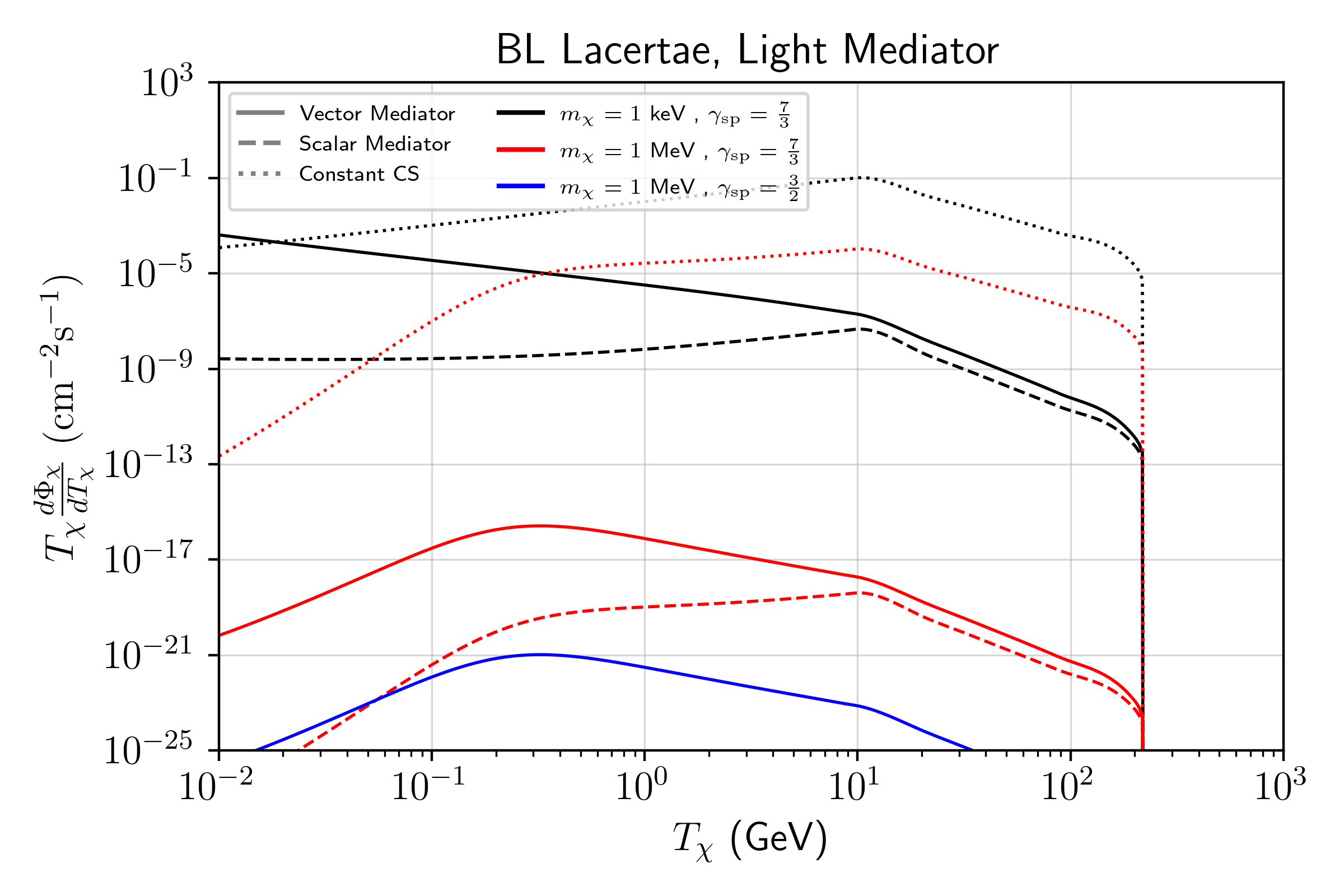}
\caption{BBDM flux for light mediator scenario}
\label{fig:lightmed_BLLac}
\end{subfigure}
\caption{Flux of DM particles, boosted by energetic electrons in the jets of BL Lacertae blazar, is plotted above, for heavy (\ref{fig:heavymed_BLLac}) and light (\ref{fig:lightmed_BLLac}) mediators.  The parameters chosen for the above plots are $\bar{\sigma}_{e\chi}=10^{-30}~\text{cm}^2$ and BMP1. The vector and the scalar mediator cases have been plotted in solid and dashed lines respectively. For comparision, DM flux for constant cross section scenario have also been plotted in dotted lines. Two DM masses have been considered, $m_\chi =1~\text{keV}$ (plotted in black) and $m_\chi = 1~\text{MeV}$ (plotted in red) for DM density Profile 1 (i.e. $\gamma_{\text{sp}} = 7/3$). To avoid overcrowding, only vector mediator case is considered for Profile 2 (i.e. $\gamma_{\text{sp}} = 3/2$), and BBDM flux is plotted (in blue) corresponding to DM mass $m_\chi = 1~\text{MeV}$. Clearly, Profile 2 yields a smaller BBDM flux as compared to Profile 1.}
\label{fig:BBDM Flux_BLLac}
\end{figure}

Inspite of this, we note in Figs.~\ref{fig:BBDM Flux_TXS} and \ref{fig:BBDM Flux_BLLac} that the flux of DM particles boosted by TXS 0506+056 is larger than the BBDM flux of BL Lacertae, for more energetic DM particles ($T_\chi \gtrsim 10~\text{GeV}$). This is because, the kinetic energy range of the electron responsible for boosting the DM particle to energies greater than $10~\text{GeV}$ is roughly $T_e \gtrsim 10~\text{GeV}$. For this energy range, the electron spectrum in TXS blazar is larger than that of BL Lac (Fig.~\ref{fig:Blazar_jet_flux}). As a result of this, we expect stronger bounds to arise from TXS blazar.

The obtained boosted DM flux will yield the following rate of electron recoil events in {\sc{Super-K}}
\begin{eqnarray}
\frac{dR}{dE_R}=\aleph \int_{T_{\chi}^{\text{min}}}^{\infty} dT_{\chi}  \frac{d\phi_{\chi}}{dT_{\chi}}  \frac{d\sigma_{\chi e}}{dE_R}
\label{eq:rate}
\end{eqnarray}

where $\aleph=7.5 \times 10^{33}$ is the effective number of target electrons in {\sc{Super-K}}~\cite{Super-Kamiokande:2017dch}, and $d\sigma_{\chi e}/dE_R$ is the differential DM-target electron interaction cross section, given by 
\begin{equation}
\frac{d \sigma_{\chi e}}{d E_{R}} = \frac{|\mathcal{M}|^{2}}{16 \pi s_\chi} \frac{1}{E_R^{\max }}
\label{eqn:diffxsecER}
\end{equation}
where $s_\chi$ is centre of momentum energy for the DM-target electron collision which can be obtained from Eqn.~\eqref{eqn:se} under the substitution : $m_\chi\leftrightarrow m_e$ and $T_{\text{e}}\rightarrow T_\chi$ . $E_R ^{\text{max}}$ is the maximum possible recoil in the detector, that can be imparted by a DM particle with kinetic energy $T_\chi$, and can be obtained from Eqn.~\eqref{eqn:Tchimax} with the appropriate substitutions mentioned before.

To get total number of expected recoil events ($N_{e\chi}$) in a certain energy bin, Eqn.~\eqref{eq:rate} needs to be integrated over $E_R$, as follows
\begin{eqnarray}
N_{e\chi} = \aleph~T_{\text{exp}} \int_{E_{R,\text{min}}} ^{E_{R,\text{max}}} dE_R \int_{T_{\chi}^{\text{min}}}^{\infty} dT_{\chi} \frac{d\phi_{\chi}}{dT_{\chi}} \frac{d\sigma_{\chi e}}{dE_R}
\label{eq:No_of_events}
\end{eqnarray}
where $T_{\text{exp}}=2628.1$ days is the exposure time~\cite{Super-Kamiokande:2017dch}, and $\left[ E_{R,\text{min}},E_{R,\text{max}} \right]$ is the recoil energy range of each bin.

\section{Simplified DM Model}
We assume that a fermionic DM particle $\chi$, of mass $m_\chi$, only interacts with electrons. This scenario is possible in several leptophilic particle DM models~\cite{Pospelov:2007mp,Batell:2009di,Chu:2011be,Alves:2013tqa,Izaguirre:2013uxa,Izaguirre:2015yja,Krnjaic:2015mbs,Izaguirre:2017bqb,Harigaya:2019shz,Bernreuther:2020koj}. Additionally, the electron-DM interaction is mediated by a scalar or vector particle, given as
\begin{eqnarray}
 \mathcal{L} &=& g_{\chi \phi} \phi \bar{\chi}{\chi} + g_{e \phi} \phi \bar{e} e \quad \,\text{or}\\ 
  &=& g_{\chi A^\prime} A^\prime_\mu \bar{\chi}\gamma^\mu{\chi} + g_{e A^\prime} A^{\prime}_\mu \bar{e}\gamma^\mu e
\end{eqnarray}
For simplicity, we will drop $A'$ and $\phi$ from subscripts in the coupling constants such that $g_\chi$ ($g_e$) is the coupling constant of the dark mediator to the DM particle  (electron). Next, we provide the differential cross-section for different operators and inspect the effect of the Lorentz structure. For that, we define the following quantities : 
\begin{eqnarray}
\mathbb{M}^2 &=&\frac{16 g_{e}^{2} g_{\chi}^{2} m_{e}^{2} m_{\chi}^{2}}{\left(q_{\mathrm{ref}}^{2}-m_{i}^{2}\right)^{2}}\\
\bar{\sigma}_{e \chi} &=& \frac{\mu_{\chi e}^2}{16 \pi m_e^2 m_\chi^2} \mathbb{M}^2
\end{eqnarray}
where $q_{\rm ref} = \alpha m_e$ is the reference momentum transferred. Here $m_i$ is the mass of the dark mediator ($i=A',\phi$ for vector, scalar mediator) and $\mu_{e\chi}$ is the reduced mass of the DM-electron system. We also define a form factor,
\begin{equation}
F_{\rm DM}^2 (q^2) = |\mathcal{M}|^{2} / \mathbb{M}^2
\label{eqn:fdmsq}
\end{equation}
This factor contains the energy dependence arising in the differential cross section $d \sigma_{\chi e}/ d T_{\chi}$ due to the blazar jet electrons boosting the DM particles and the Lorentz structure of the interaction. The explicit form of $F_{\rm DM}$ depends on the model of DM and mediator considered.

A similar form factor, $F_{\rm rec}$, contains the energy dependence in the differential cross section $d \sigma_{\chi e}/ d E_{R}$ arising due to the interaction of relativistic DM particles with the electrons in {\sc{Super-K}}, and can be obtained from the form factor $F_{\rm DM}$ of Eqn.~\eqref{eqn:fdmsq} by making the substitutions : $m_e \leftrightarrow m_\chi$, $T_\chi \rightarrow E_R$ and $T_e \rightarrow T_\chi$.

Hence the differential cross sections, $d \sigma_{\chi e}/ d T_{\chi}$ and $d \sigma_{\chi e}/ d E_{R}$, relevant in the DM-blazar jet electron scattering and DM scattering at the detector end respectively, are given by :
\begin{eqnarray}
\frac{d\sigma_{\chi e}}{dT_{\chi}}= \bar{\sigma}_{e \chi} \frac{m_e ^2 m_{\chi} ^2}{\mu _{e \chi} ^2} \frac{ F_\text{DM} ^2 (q^2)}{s_{\text{CR}} T_{\chi} ^{\text{max}}} 
\label{eqn:dsig/dT}
\end{eqnarray}
and,
\begin{eqnarray}
\frac{d\sigma_{\chi e}}{dE_{R}}= \bar{\sigma}_{e \chi} \frac{m_e ^2 m_{\chi} ^2}{\mu _{e \chi} ^2} \frac{ F_\text{rec} ^2 (q^2)}{s_\chi E_{R} ^{\text{max}}} 
\label{eqn:dsig/dER}
\end{eqnarray}

Under the energy independent approximation for the cross section, the differential cross section would simply be :
\begin{eqnarray}
\frac{d\sigma_{\chi e}}{dT_{\chi}}=\frac{ \bar{\sigma}_{e \chi}}{T_{\chi} ^{\text{max}}} ~,~ \frac{d\sigma_{\chi e}}{dE_R}=\frac{ \bar{\sigma}_{e \chi}}{E_R ^{\text{max}}}
\end{eqnarray}

\subsection{Scalar Mediator}
Considering a scalar mediator (denoted as $\phi$), one can calculate $F_{\rm DM}^2$ for the interaction between electrons in blazar jets and non-relativistic DM, using Eqn.~\eqref{eqn:fdmsq} to obtain
\begin{equation}
F_{\mathrm{DM}}^{2}(q)=\frac{\left(q_{\mathrm{ref}}^{2}-m_{\phi}^{2}\right)^{2}}{\left(q^{2}-m_{\phi}^{2}\right)^{2}} \frac{\left(2 m_{\chi}+T_{\chi}\right)\left(2 m_{e}^{2}+m_{\chi} T_{\chi}\right)}{4 m_{\chi} m_{e}^{2}}
\label{eqn:ff-scalar}
\end{equation}

The differential cross section ($d\sigma_{\chi e}/dT_{\chi}$) w.r.t. the DM energy ($T_{\chi}$), is :
\begin{eqnarray}
\frac{d\sigma_{\chi e}}{dT_{\chi}} &=& \tilde{\sigma}_{e \chi}  \frac{(q_{\mathrm{ref}}^{2}-m_{\phi}^{2})^{2}}{(q^{2}-m_{\phi}^{2})^{2}} 
\left\{ \frac{m_{\chi}}{4 \mu _{e \chi} ^2}\right.
\nonumber \\
&&\left.  \frac{\left(2 m_{\chi}+T_{\chi}\right)\left(2 m_{e}^{2}+m_{\chi} T_{\chi}\right)}{ s_{\text{CR}} T_{\chi} ^{\text{max}}} \right\}
\label{eqn:sigma-scalar}
\end{eqnarray}

The form factor $F_{\text{rec}}$ and the differential cross-section w.r.t. the recoil energy of the detector ($d\sigma_ {\chi e}/dE_R$) are obtained from Eqn.~\eqref{eqn:ff-scalar} and Eqn.~\eqref{eqn:sigma-scalar} by performing the substitutions prescribed in the previous section, viz. $m_e \leftrightarrow m_\chi$, $T_\chi \rightarrow E_R$, $T_e \rightarrow T_\chi$, $s_e \rightarrow s_{\chi}$ .

\subsection{Vector Mediator}
Using a similar treatment for the vector mediator (denoted by $A'$), we find that
\begin{eqnarray}
F_{\mathrm{DM}}^{2}(q^2)&=&\frac{\left(q_{\mathrm{ref}}^{2}-m_{A'}^{2}\right)^{2}}{\left(q^{2}-m_{A'}^{2}\right)^{2}} \frac{1}{2 m_{\chi} m_{e}^{2}} \left( 2 m_{\chi}\left(m_{e}+T_{e}\right)^{2}-\right. \nn \\ 
&& \left. T_{\chi}\left\{\left(m_{e}+m_{\chi}\right)^{2}+2 m_{\chi} T_{e}\right\}+m_{\chi} T_{\chi}^{2}\right)
\end{eqnarray}
and,
\begin{eqnarray}
\frac{d\sigma_{\chi e}}{dT_{\chi}} &=& \bar{\sigma}_{e \chi} \frac{\left(q_{\mathrm{ref}}^{2}-m_{A'}^{2}\right)^{2}}{\left(q^{2}-m_{A'}^{2}\right)^{2}} \frac{m_{\chi}}{2 \mu _{e \chi} ^2 s_{\text{CR}} T_{\chi} ^{\text{max}}}\left\{ 2 m_{\chi}(m_{e}+ T_{e})^{2}  \right.
\nonumber \\
&&\left. - T_{\chi} \{ (m_{e}+m_{\chi})^{2}+2 m_{\chi} T_{e}\} + m_{\chi} T_{\chi}^{2}\right\}
\label{eqn:sigma-vector}
\end{eqnarray}

\section{Results}
The effect of including energy dependence in DM-electron interaction can be seen from the BBDM flux plots, given in Figs.~\ref{fig:BBDM Flux_TXS} and \ref{fig:BBDM Flux_BLLac}. Profile 1 of DM density clearly gives larger BBDM flux as compared to Profile 2, as expected from larger DM spike for profile 1 shown in  Fig.~\ref{fig:DMDensity} in Section III. For DM to register event at {\sc{Super-K}}, kinetic energies greater than $\sim 0.1~\text{GeV}$ are relevant. In this energy range, the heavy mediator scenario gives much larger BBDM flux as compared to the constant cross section case, while on the other hand, the light mediator regime yields a much smaller BBDM flux. From this, we expect the exclusion limit on DM-electron interaction, arising from light mediator regime, to be extremely weak. Since the vector mediator case gives slightly larger BBDM flux as compared to the scalar mediator scenario, we hope for moderately better bounds from vector mediators. Also, since for any given DM Profile or BMP, the BBDM flux is larger for smaller mass DM particles, we can expect bounds to grow stronger for lighter DM particle. Finally, the BBDM flux plots terminate at a certain value of $T_{\chi}$ because the blazar jet electrons boosting the DM particles have an upper cutoff on their energies (for TXS 0506+056 jets, $T_{e,\text{jet}}^{\text{max}} \sim 260~\text{GeV}$ and for BL Lacertae jets, $T_{e,\text{jet}}^{\text{max}} \sim 225~\text{GeV}$).

Taking into account the signal efficiency of each recoil bin ($\epsilon_{\text{sig}}$), the exclusion limit on $\bar{\sigma}_{e \chi}$ is obtained by
\begin{eqnarray}
N_{e\chi} \epsilon_{\text{sig}} < N_{\text{B}},
\label{eq:Ne-NB}
\end{eqnarray}

where $N_{e\chi}$, obtained from Eqn.~\eqref{eq:No_of_events}, is the number of expected recoil events arising out of collision of target electron with DM particles boosted by the blazars. $N_{\text{B}}$ ($B= {\rm TXS, BL}$ for TXS 0506+056 and BL Lacertae) are the $95\%$ CL upper limits on number of events from the blazars. 

Three energy bins were considered in the analysis released by {\sc{Super-K}} collaboration \cite{Super-Kamiokande:2017dch}. The total number of events, the Monte Carlo simulation of the background, signal efficiency and spatial distribution of events were provided for each bin. One can use this data to select signals from a certain ``searching cone" in the direction of the blazar. This removes the majority of the background from the data, increasing sensitivity. The selected signal is then used in the standard Poisson method~\cite{ParticleDataGroup:2020ssz}  to yield $95\%$ CL upper limit on expected number of events ($N_{\text{B}}$) for each of the three bins. This analysis was performed by the authors of Ref.~\cite{Granelli:2022ysi}, and we use their results (i.e. $N_{\text{B}}$), summarised in Table \ref{table:superK_events}. (For details of the analysis, see \cite{Super-Kamiokande:2017dch,Ema:2018bih,Cappiello:2019qsw,Granelli:2022ysi}). This gives us all the numbers relevant to finding exclusion limits on $\bar{\sigma}_{e \chi}$ using Eqn.~\eqref{eq:Ne-NB}. 

\begin{table}[htb]
\resizebox {\columnwidth}{!}{
\begin{tabularx} {0.4\textwidth}{ 
   >{\centering\arraybackslash}X 
   >{\centering\arraybackslash}X 
   >{\centering\arraybackslash}X 
   >{\centering\arraybackslash}X
   >{\centering\arraybackslash}X  }
\hline 
Bins & $E_R(\text{GeV})$ & $\epsilon_{\text{sig}}$ & $N_{\text{TXS}}$ & $N_{\text{BL}}$ \\
\hline \hline
Bin 1 & $(0.1,~1.33)$ & $93.0\%$ & $19.39$ & $17.27$ \\
Bin 2 & $(1.33,~20)$ & $91.3\%$ & $3.42$ & $6.27$ \\
Bin 3 & $(20,~10^3)$ & $81.1\%$ & $2.98$ & $2.98$ \\
\end{tabularx}
}
\caption{Signal efficiency ($\epsilon_{\text{sig}}$) and $95\%$ CL upper limits on number of events ($N_{\text{TXS}}$ and $N_{\text{BL}}$) from the blazars, provided for the three recoil bins of {\sc{Super-K}}.}
\label{table:superK_events}
\end{table}

{\sc{Super-K}} is located deep underground to reduce background. As a result, the DM flux entering the detector is significantly attenuated, primarily as a result of its interaction with electrons on the Earth's surface, and this gives rise to the attenuation bound in the exclusion plot. We provide an approximation of the attenuation bound, which is the cross section for which the DM particle with $T_\chi \sim 10\,{\rm GeV}$ can impart the threshold recoil energy in the detector. For this, we solve the following equation to calculate the energy ($T_r$) lost by the DM
\begin{eqnarray}
\label{eq:Tchiz}
\frac{dT_\chi}{dx} = -\sum_T n_T \int_0 ^{T_r ^{\text{max}}} \frac{d\sigma}{dT_r}T_rdT_r
\end{eqnarray}
and estimate $\bar{\sigma}_{e \chi}$ so that kinetic energy of the DM particle at depth $z$, denoted by $T_\chi ^z$, is the detector threshold $E_{\rm th}$ (we consider $E_{\text{th}}=100~{\text{MeV}}$ corresponding to Bin 1), for an initial kinetic energy $T_{\chi,{\rm{in}}} \sim 10~\text{GeV}$. The area bounded by the attenuation bound and the exclusion bound is ruled out by our analysis. In this work, we limit ourselves to elastic scattering and ignore backscattering of light DM particles into the atmosphere. Note that the attenuation limits exist only for the heavy mediators and this may also vary once a more elaborate study is performed, which we leave for future work.

\begin{figure}[!h]
\begin{subfigure}{\linewidth}
\hspace*{-6 mm}
\includegraphics[scale=0.6]{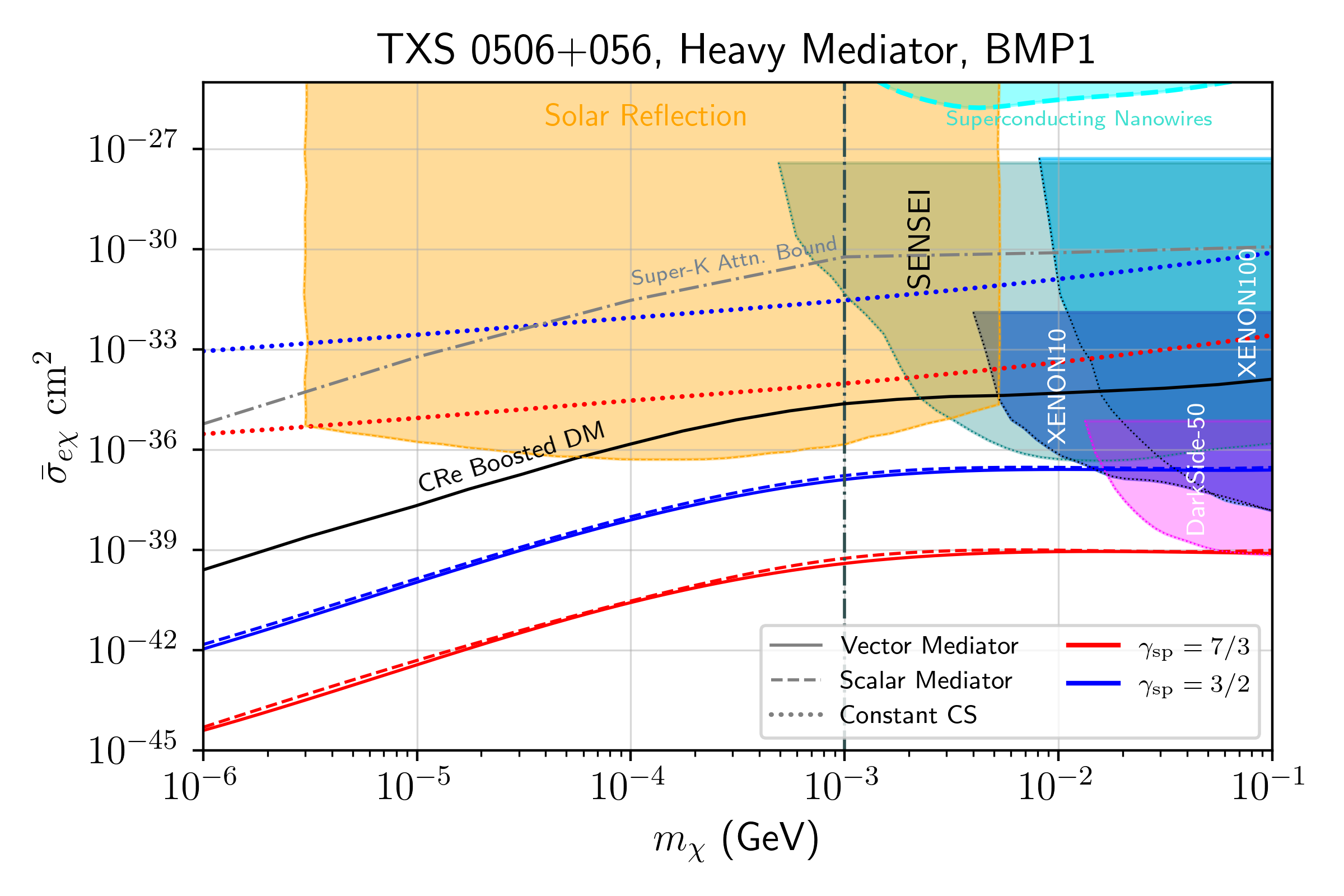}
\caption{Exclusion Bound for TXS 0506+056}
\label{fig:TXS_Excl_HeavyMed_BMP1}
\end{subfigure}
\begin{subfigure}{\linewidth}
\hspace*{-6 mm}
\includegraphics[scale=0.6]{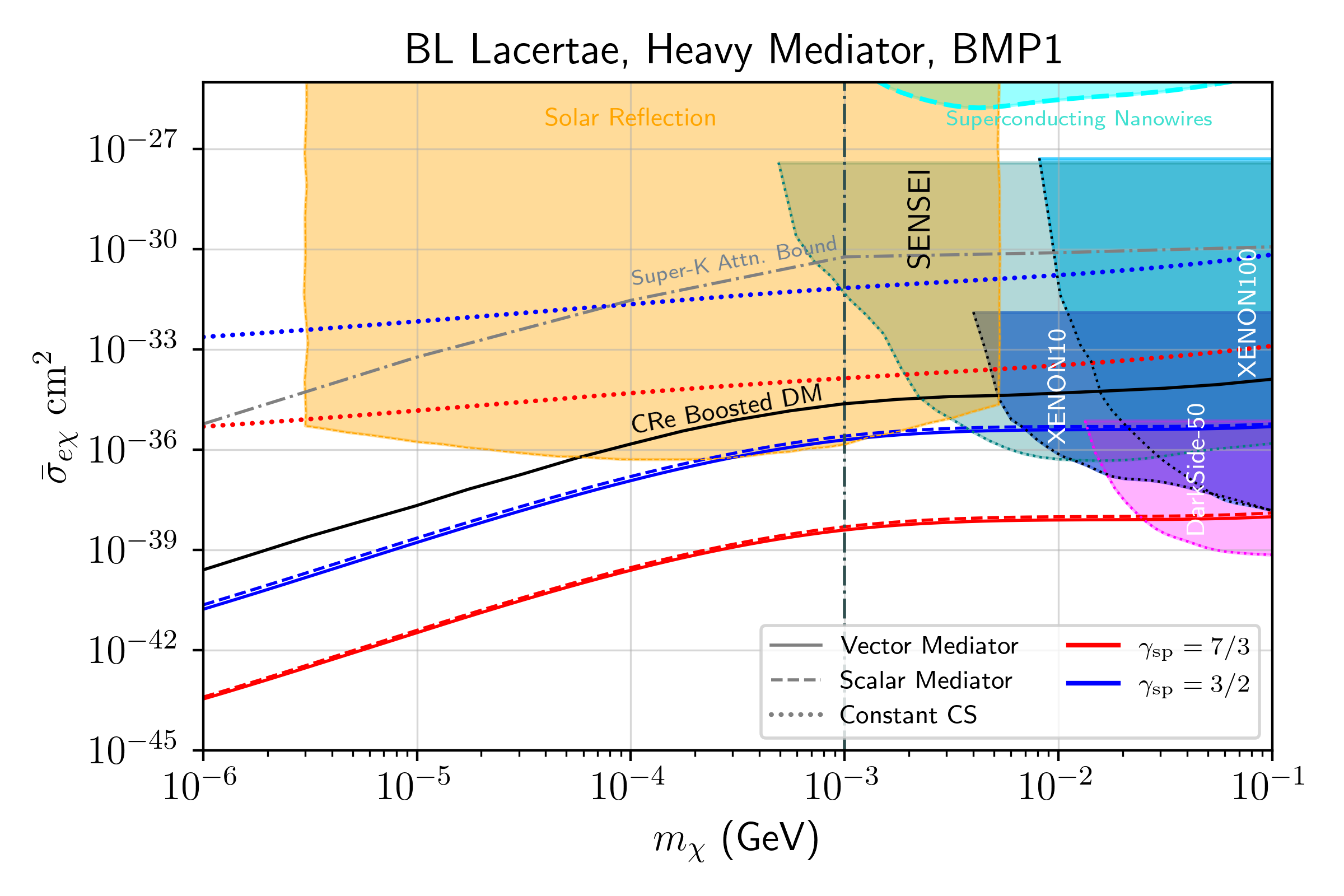}
\caption{Exclusion Bound for BL Lacertae}
\label{fig:BLLac_Excl_HeavyMed_BMP1}
\end{subfigure}
\caption{The exclusion bound is plotted for the blazars TXS 0506+056 (\ref{fig:TXS_Excl_HeavyMed_BMP1}) and BL Lacertae (\ref{fig:BLLac_Excl_HeavyMed_BMP1}), corresponding to BMP1. The cases considered are heavy vector mediator (plotted in solid lines), heavy scalar mediator (plotted in dashed lines) and constant cross section case (plotted in dotted lines). The various DM density profiles considered are Profile 1 (in red) and Profile 2 (in blue). The direct detection bounds from {\sc Xenon10,  Xenon100, SENSEI}~\cite{Essig:2012yx,Essig:2017kqs,SENSEI:2020dpa} and {\sc DarkSide-50}~\cite{DarkSide-50:2022hin} are also plotted. {The BBN limits are shown by dotted-dashed curve.}
The bound arising due to DM attenuation is also given for heavy mediator scenario (plotted in grey). The area bounded by the attenuation bound and the exclusion bound is ruled out by our analysis. Exclusion limit from Superconducting Nanowires is provided in cyan color. Constraint due to solar reflection of DM \cite{An:2017ojc,Cao:2020bwd} is shown in amber color. The exclusion bound given by Cosmic Ray electron (CRe) boosted DM~\cite{Bardhan:2022ywd} is plotted in black (See text for more details).
}
\label{fig:ExclBound_BMP1}
\end{figure} 

\begin{figure}[!h]
\begin{subfigure}{\linewidth}
\hspace*{-6 mm}
\includegraphics[scale=0.6]{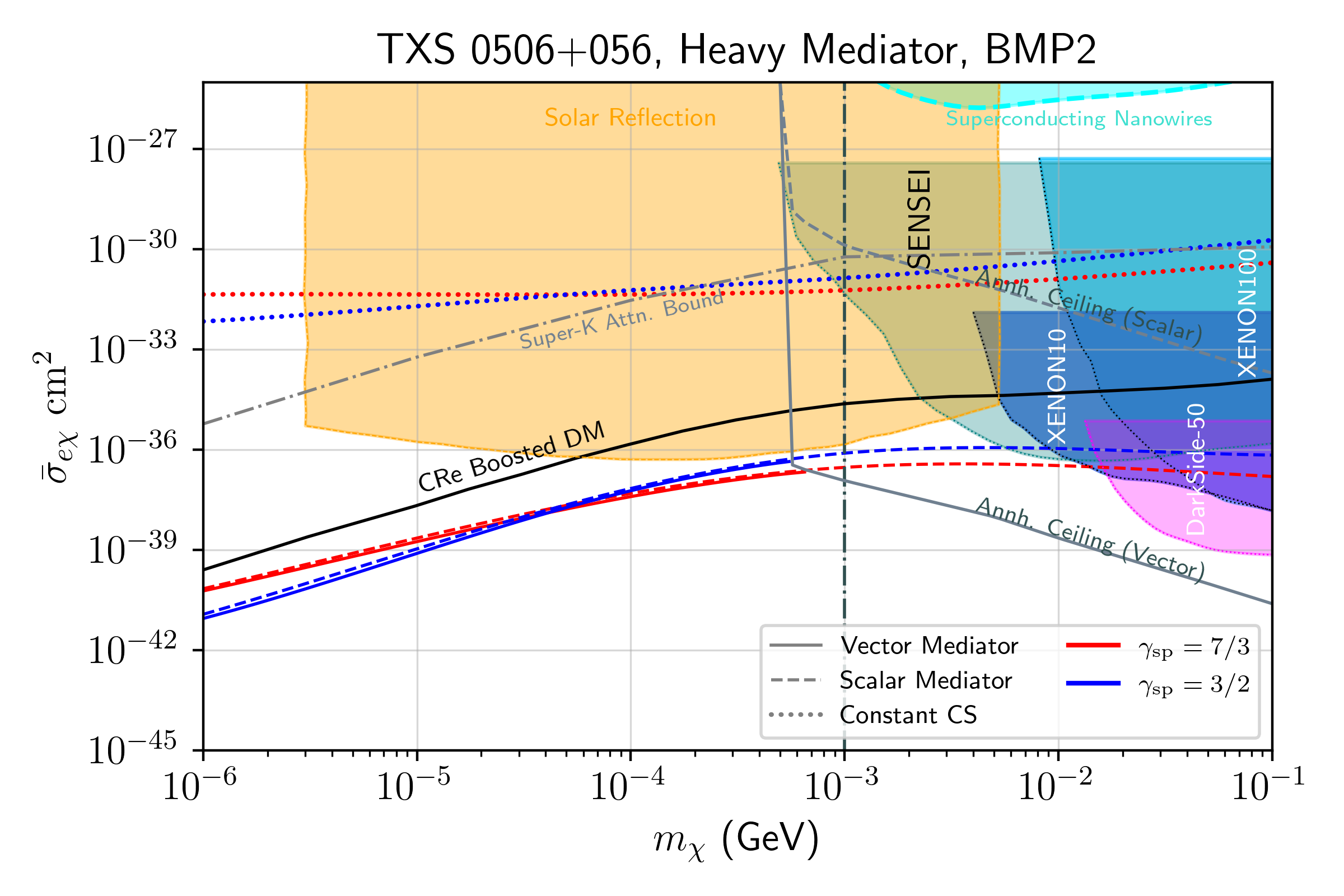}
\caption{Exclusion Bound for TXS 0506+056}
\label{fig:TXS_Excl_HeavyMed_BMP2}
\end{subfigure}
\begin{subfigure}{\linewidth}
\hspace*{-6 mm}
\includegraphics[scale=0.6]{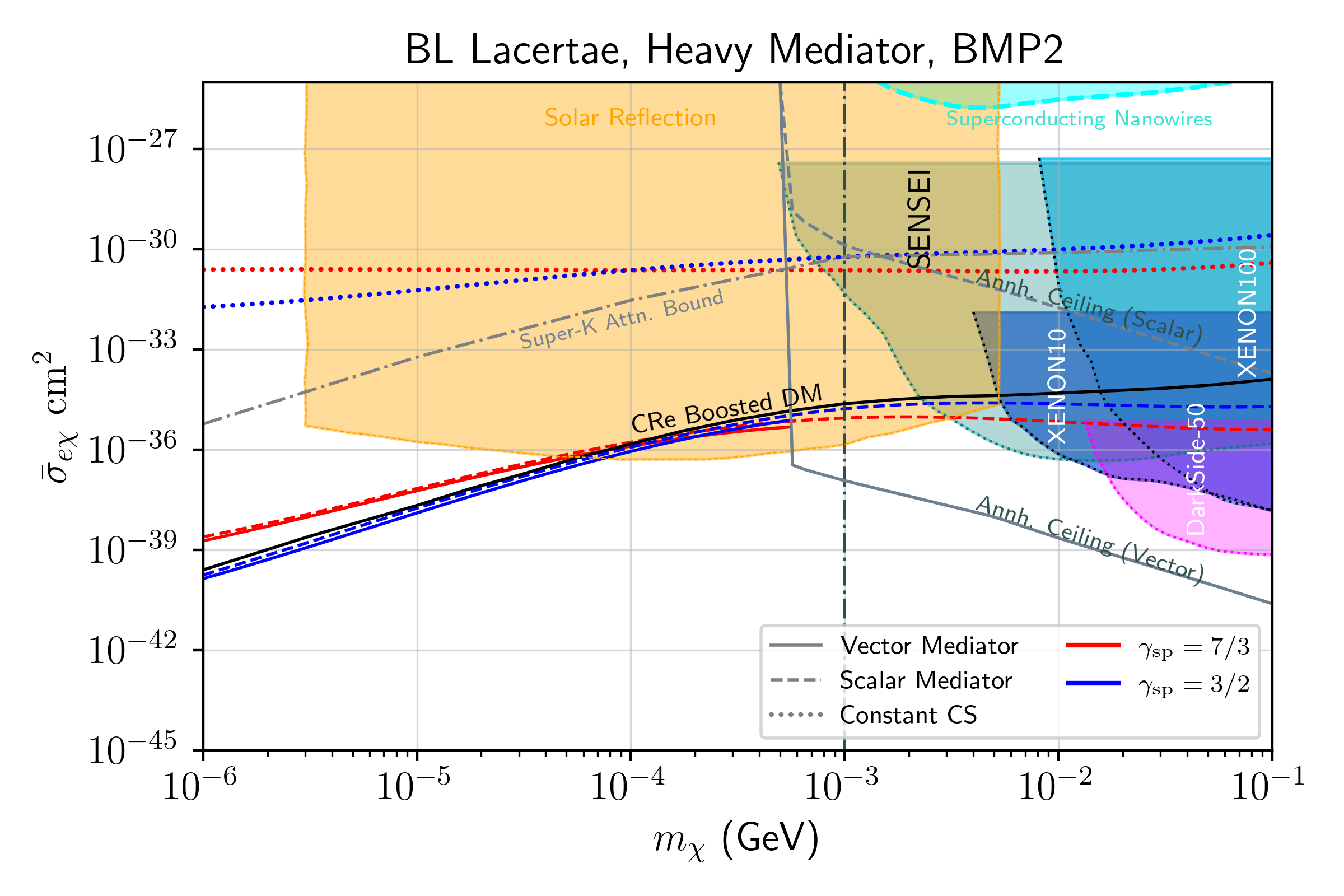}
\caption{Exclusion Bound for BL Lacertae}
\label{fig:BLLac_Excl_HeavyMed_BMP2}
\end{subfigure}
\caption{The exclusion bound is plotted for the blazars TXS 0506+056 (\ref{fig:TXS_Excl_HeavyMed_BMP2}) and BL Lacertae (\ref{fig:BLLac_Excl_HeavyMed_BMP2}), corresponding to BMP2. The cases considered are heavy vector mediator (plotted in solid lines), heavy scalar mediator (plotted in dashed lines) and constant cross section case (plotted in dotted lines). The various DM density profiles considered are Profile 1 (in red) and Profile 2 (in blue). The direct detection bounds from {\sc Xenon10, Xenon100, SENSEI}~\cite{Essig:2012yx,Essig:2017kqs,SENSEI:2020dpa} and {\sc DarkSide-50}~\cite{DarkSide-50:2022hin} are also plotted. The bound arising due to DM attenuation is also given for heavy mediator scenario (plotted in grey). The area bounded by the attenuation bound and the exclusion bound is ruled out by our analysis. {The BBN limits are shown by dotted-dashed curve.} Exclusion limit from Superconducting Nanowires is provided in cyan color. Constraint due to solar reflection of DM \cite{An:2017ojc,Cao:2020bwd} is shown in amber color. The exclusion bound given by Cosmic Ray electron (CRe) boosted DM~\cite{Bardhan:2022ywd} is plotted in black (See text for more details). 
% { Note that BBN bounds may be relevant for $m_\chi\,<\,1\,{\rm MeV}$, but a full understanding of the dark matter model is required to properly assess these bounds.}
}
\label{fig:ExclBound_BMP2}
\end{figure}

\begin{figure}[!h]
\hspace*{-6 mm}
\includegraphics[scale=0.6]{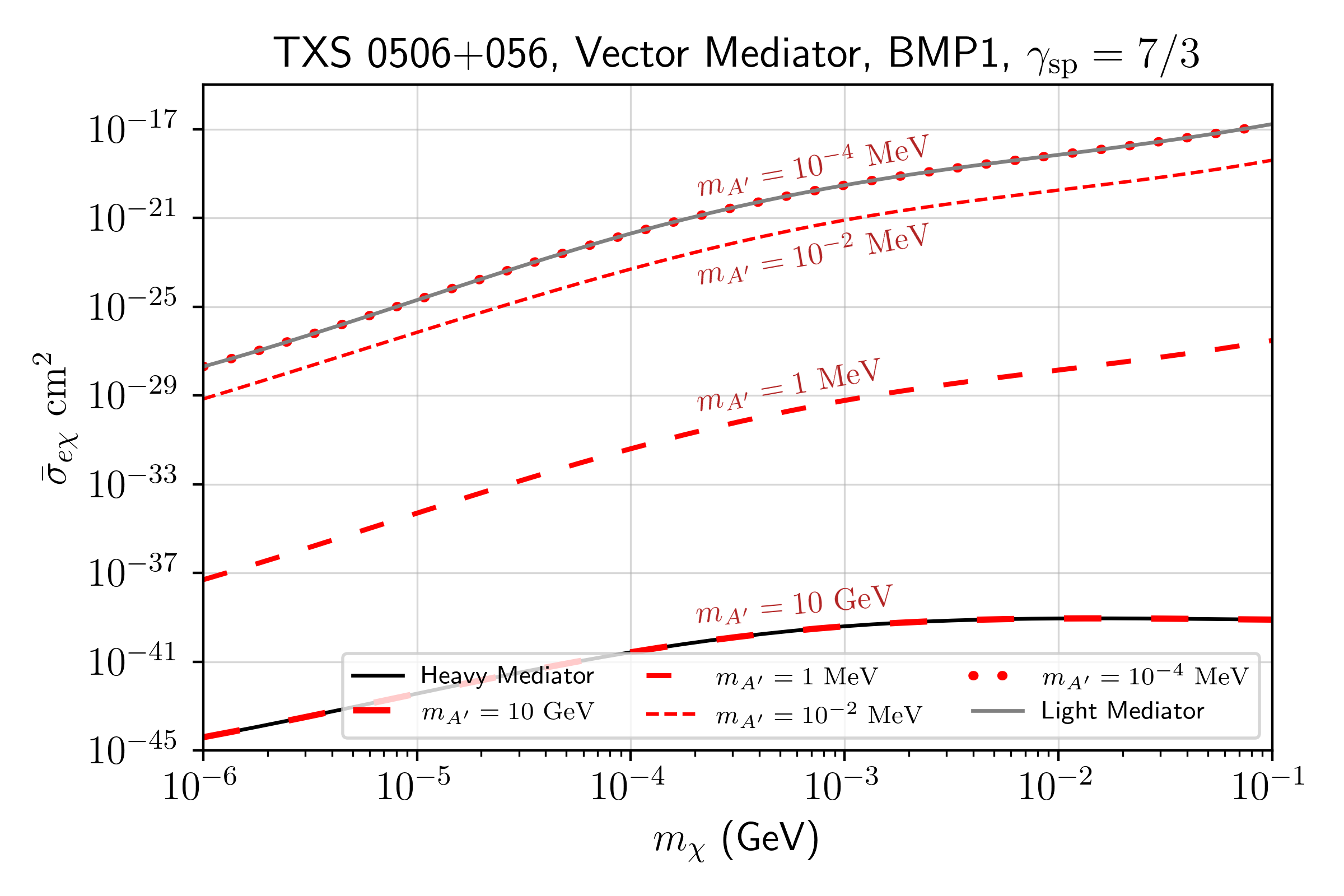}
\caption{The exclusion bound is plotted for the blazar TXS 0506+056 for various mediator masses, corresponding to the vector mediator scenario. The mediator masses chosen are $10~\text{GeV}$, $1~\text{MeV}$, $10^{-2}~\text{MeV}$ and $10^{-4}~\text{MeV}$, all plotted in different linestyles. The bound for heavy (in black) and light (in grey) mediator regime is also plotted. The profiles chosen are DM density Profile 1, BMP1.}
\label{fig:ExclBound_TXS}
\end{figure} 

The exclusion bound arising from {\sc{Super-K}} data is shown in Figs.~\ref{fig:ExclBound_BMP1} and \ref{fig:ExclBound_BMP2} in the heavy mediator regime for scalar and vector operators. Taking energy dependence into account, the exclusion bound is significantly different compared to the bound obtained from constant cross section assumption. BMP1 sets a stronger bound as compared to BMP2, and DM density Profile 1 yields a better bound as compared to Profile 2. This is in agreement with what we expected from the density profiles (Fig.~\ref{fig:DMDensity}) in Section III and the BBDM flux plots (Figs.~\ref{fig:heavymed_TXS} and \ref{fig:heavymed_BLLac}) in Section IV.

Amongst the three recoil bins in {\sc{Super-K}}, the strongest bound is set by Bin 3 ( Bin 2 ) for heavy mediator (constant cross section) cases. This result can be explained from the BBDM flux plots for TXS (Fig.~\ref{fig:heavymed_TXS}), by observing that the BBDM flux for heavy mediator is largest for $T_\chi \sim (20~\text{GeV},10^3~\text{GeV})$, which is nearly the DM energy range relevant to produce recoil in the third bin of {\sc{Super-K}}. Similarly, for constant cross section case, the BBDM flux is largest for $T_\chi \sim (1~\text{GeV},20~\text{GeV})$, which is roughly the DM energy range relevant to the second recoil energy bin of {\sc{Super-K}}. For BL Lacertae, even though the BBDM flux is largest $T_\chi \sim (1~\text{GeV},20~\text{GeV})$ , which is the DM energy range relevant for Bin 2 (Fig.~\ref{fig:heavymed_BLLac}), the event rate is largest for Bin 3 due to the larger size of the Bin. We also note that the constraints from TXS 0506+056 is stronger than constraints from BL Lacertae, which is what we expected from Fig.~\ref{fig:Blazar_jet_flux} and the discussion in Section IV.

In the exclusion bound plots (Figs.~\ref{fig:ExclBound_BMP1} and \ref{fig:ExclBound_BMP2}), we plot the most stringent limit on $\bar{\sigma}_{e\chi}$ coming from the three bins. Note that the bounds arising from scalar and vector mediators are almost same. The reason for this can be understood from Figs.~\ref{fig:BBDM Flux_TXS} and \ref{fig:BBDM Flux_BLLac}, where we see that the BBDM flux for DM energies relevant to recoil Bin 3 of {\sc{Super-K}} differs by $\sim \mathcal{O}(10)$ for the two operators, which results in very little difference in the exclusion limits. However, the limit coming from the three different bins can differ by $\sim$ 2 or 3 orders of magnitude for certain $m_{\chi}$, so a combined analysis of the three bins might change the bounds. We, however, leave out such an analysis from this work.

Since ``heavy" and ``light" mediator regimes are the convenient extremes of the actual DM model, the true exclusion bound would lie somewhere in between the bound set by these two regimes. Thus we compare the exclusion bound corresponding to various masses of the vector mediator for Profile 1 and BMP1 in Fig.~\ref{fig:ExclBound_TXS}. A similar comparison and scaling exists for Profile 2 and BMP2.  Clearly, a mediator of mass $10~{\text{GeV}}$ corresponds to the heavy regime, and a mediator of mass $10^{-4}~{\text{MeV}}$ reproduces the exclusion limit set by the light mediator regime. 

It should be noted that for $m_{\chi}\geq\,m_e$ the scattering cross-section and consequently the annihilation rate to $e^-e^+$ increases as we move to heavier DM particles in Fig.~\ref{fig:ExclBound_BMP2} (for $m_{\chi}\leq\,m_e$, $\sigma_{\rm ann}^{e^-e^+}=0$). The DM annihilation rate to $e^-e^+$ for scalar mediator is given by
\begin{eqnarray}
\sigma_\phi =\frac{1}{16 \pi} \frac{(g_e g_\chi)^2}{(q^2 - m_\phi^2)^2} \sqrt{\frac{s-4m_e^2}{s-4m_\chi^2}} \frac{(s-4m_e^2)(s-4m_\chi^2)}{s}
\end{eqnarray}
while for the vector mediator case, the annihilation rate is
\begin{eqnarray}
\sigma_{A'} =\frac{1}{12 \pi} \frac{(g_e g_\chi)^2}{(q^2 - m_{A'} ^2)^2} \sqrt{\frac{s-4m_e^2}{s-4m_\chi^2}} [ s + \nonumber \\ 
 2(m_\chi ^2+m_e^2) + \frac{4m_\chi ^2 m_e^2}{s}]
\end{eqnarray}
where $s$ is centre of momentum frame energy and $v_\chi \approx 10^{-3}$ is DM velocity near centre of blazar. As a result, for large enough cross sections, DM annihilation rate can exceed $3\,\times\,10^{-26}{\rm cm}^3s^{-1}$, contradicting the rate assumed for BMP2. We refer to these cross sections as the annihilation ceiling and they are shown by gray curves in Fig.~\ref{fig:ExclBound_BMP2}. For vector interaction, the annihilation ceiling established for a 100 GeV mediator is lower than the bound set by our work, thereby rendering the exclusion limits irrelevant. Hence, our exclusion limits for vector mediator are shown only for $m_{\chi}<m_e$. However, a significant parameter space remains constrained by our exclusion bounds for scalar interaction. There is no annihilation ceiling for BMP1 because it is presumed that annihilation is prohibited in BMP1 (for instance, in the case of asymmetric DM~\cite{Petraki:2013wwa}). 

Apart from blazars jets, Cosmic Ray electrons (CRe) provide yet another environment to produce boosted DM particles. Exclusion bound from CRe boosted DM, using {\sc{Super-K}} data, is plotted alongwith our bounds in Figs.~\ref{fig:ExclBound_BMP1} and \ref{fig:ExclBound_BMP2}. Furthermore, Refs.~\cite{Hochberg:2019cyy, Hochberg:2021yud} propose a DM detection device with extremely low recoil trigger made using Superconducting Nanowires. The best bounds from such a prototype device is also shown. Currently our results are much stronger, but proposed devices with materials like NbN and Al might give better exclusions in the near future. Constraints from other direct detection experiments, such as {{\sc{Xenon10, Xenon100, SENSEI}}~\cite{Essig:2012yx,Essig:2017kqs,SENSEI:2020dpa} and {\sc DarkSide-50}~\cite{DarkSide-50:2022hin} are also shown. Exclusion limits from solar reflection of DM \cite{An:2017ojc,Cao:2020bwd} are important in the heavy mediator case, and are provided as well. Further, we translate our strongest bounds corresponding to blazar TXS 0506+056 onto the mass-coupling plane for $m_{A'(\phi)}\,=\,10\,m_\chi$, as shown in Fig.~\ref{fig:LabBound}, alongwith other laboratory bounds from literature. Additionally, our benchmark allows for the constraint of $g_{e}$ through the analysis of missing energy/momentum signals observed in the dark photon search conducted by the NA64 collaboration\cite{Andreev:2021fzd} and the BABAR analysis~\cite{BaBar:2017tiz}. It should be noted that when depicting these constraints for scalar mediators, we have taken a conservative approach, resulting in the elimination of a slightly larger portion of the parameter space than is strictly necessary.

\begin{figure}[!h]
\begin{subfigure}{\linewidth}
\hspace*{-6 mm}
\includegraphics[scale=0.6]{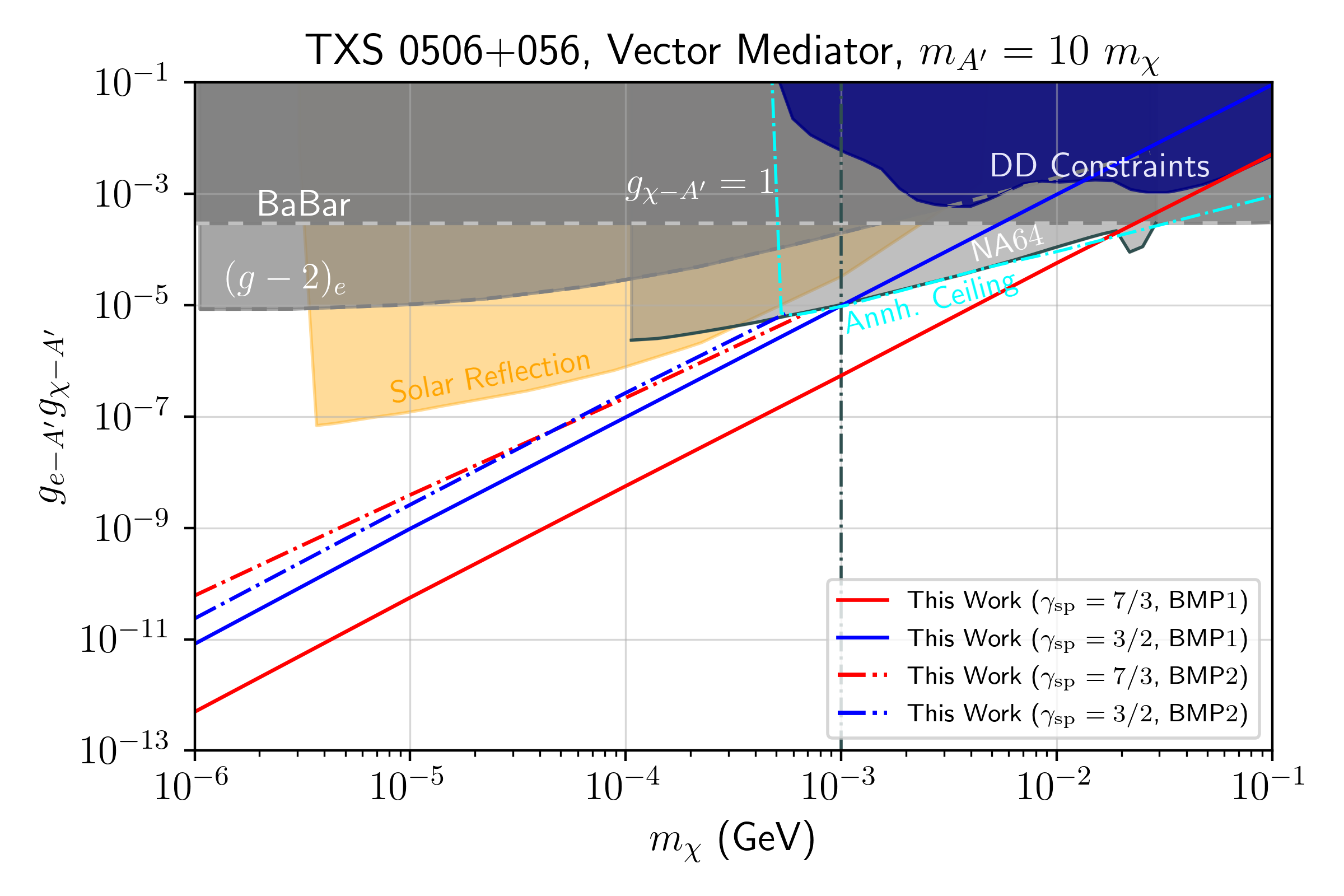}
\caption{Constraints on couplings with Vector mediator.}
\label{fig:Lab_vec}
\end{subfigure}
\begin{subfigure}{\linewidth}
\hspace*{-6 mm}
\includegraphics[scale=0.6]{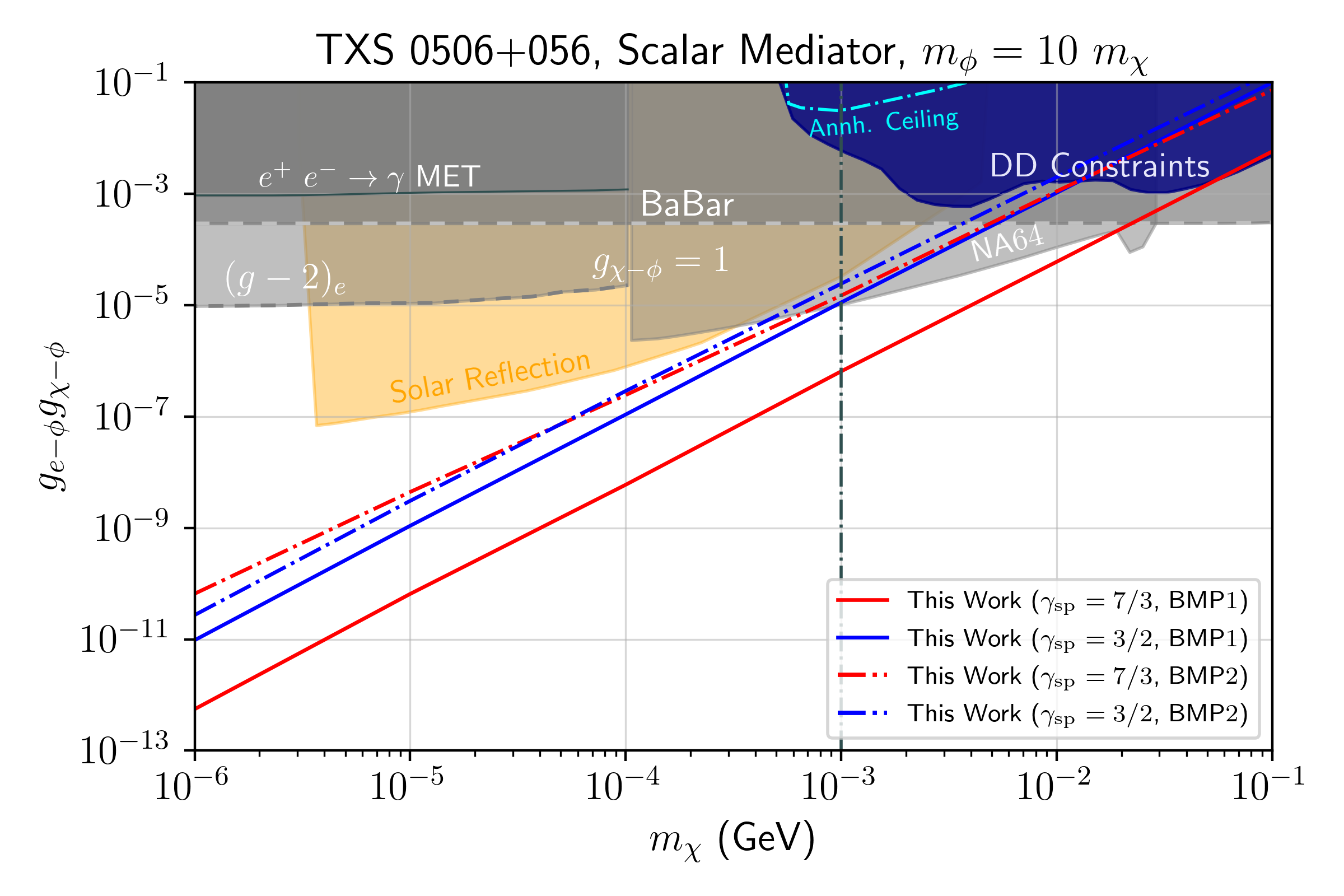}
\caption{Constraints on couplings with Scalar mediator.}
\label{fig:Lab_scal}
\end{subfigure}
\caption{The exclusion limit on $g_{e}g_{\chi}$ is plotted above for a) vector and b) scalar mediator for the blazar TXS 0506+056. The mediators are chosen to be 10 times more massive than the DM particles. Our bounds are plotted in red (Profile 1) and blue (Profile 2) curves, corresponding to BMP1 (solid lines) and BMP2 (dash-dotted lines). We also show bounds from {\sc{BaBar}}~\cite{BaBar:2017tiz}, {\sc{NA64}} \cite{Andreev:2021fzd}, Solar Reflection \cite{An:2017ojc,Cao:2020bwd}, direct detection constraints \cite{Essig:2012yx,Essig:2017kqs,SENSEI:2020dpa,DarkSide-50:2022hin}, MET searches and $(g-2)_e$ constraints \cite{Knapen:2017xzo}.}
\label{fig:LabBound}
\end{figure}

{The cosmological constraints from Big Bang Nucleosynthesis (BBN) rule out thermal DM of $m_\chi \lesssim 10{\ \rm MeV}$ stringently~\cite{Knapen:2017xzo,Ghosh:2020vti}. However, these bounds can be relaxed in DM models down to 1 MeV with couplings by switching on couplings to neutrinos besides electrons~\cite{Escudero:2018mvt}. Thus, BBN bounds are of significance when considering dark matter with a mass below 1 MeV ($m_\chi<1{\rm MeV}$). However, it is crucial to have a comprehensive understanding of the dark matter model in order to accurately evaluate these bounds.} The CMB observations similarly constrain DM annihilating to an $e^-e^+$ pair severely~\cite{Planck:2018vyg}. An elaborate dark sector associated in these models can relax these constraints, so that DM mostly annihilate to other dark sector particles~\cite{Choudhury:2020xui}. 

\section{Summary \& Outlook}
Blazars, in addition to being a key source of high energy electrons, are projected to have a DM density spike in their core due to DM accretion onto their SMBH.
Despite large uncertainties from astrophysics and the unknown annihilation properties of dark matter in the density of the succeeding DM spike, strong bounds 
on the elastic scattering cross section for DM-electron scattering have been obtained in Ref.~\cite{Granelli:2022ysi}. Here, we demonstrate how these limits change when the resulting energy dependence of the S-matrix for the associated vertex Lorentz structure is taken into consideration. We remain agnostic of the relic abundance mechanism since DM models might include an extended dark sector that has a significant impact on how much DM is there in the Universe right now. To that end, we derived limits using {\sc{Super-K}} data. We found that the constraints on such energy-dependent scattering cross sections, which mostly depend on mediator mass, are at least several orders of magnitude tighter than the current limits from Blazars in the literature for the constant cross section assumption. Though the constant cross-section is a meaningful way to explain a concept, in reality it corresponds to a small parameter space of a DM model. Our bounds are, however, weakened if the mediator mass is sufficiently small. This is because the BBDM flux is orders of magnitude less than the constant cross-section in the relevant energy bin (see fig.~\ref{fig:lightmed_TXS},\ref{fig:lightmed_BLLac}). We also studied the less cuspy profile of the DM spike and realised that the constraints on $\bar{\sigma}_{e\chi}$ from BBDM are still significant compared to Cosmic ray boosted DM. 

Another subtlety is that, in addition to relativistic electrons, blazars also contain energetic protons, which may contribute to the BBDM flux.
However, the contribution is insignificant since the coupling with the proton is loop-suppressed if there is just a tree-level interaction of DM with charged electrons. Another natural assumption, inspired by the standard model's $SU(2)_L$ gauge symmetry, is that neutrino should have the same cross section 
with DM as charged leptons~\cite{Cline:2022qld}, allowing us to compare the current $\sigma_{e\chi}\sigma_{\nu\chi}$ limits for Cosmic ray boosted DM~\cite{PhysRevD.105.103029}.
We intend to investigate this possibility in next work. 

\section*{Acknowledgements}
D.G. acknowledges support through the Ramanujan Fellowship and MATRICS Grant of the Department of Science and Technology, Government of India. D.S. has received funding from the European Union’s Horizon 2020  research and innovation programme under grant agreement No 101002846, ERC CoG “CosmoChart”. The authors also thank Arka Banerjee and Susmita Adhikari for valuable discussions. We thank Robert Mcgehee, Iason Baldes and Kalliopi Petraki for useful comments. 

\newpage
\onecolumngrid

\appendix

\section{Scaling of exclusion limits with DM density spike}
\label{appendix:A}

In this work, the normalization $\mathcal{N}_1$ in Eqn.~\eqref{eq:spikedensity} has been determined by observing that the mass of the spike is of the same order as $M_{\rm BH}$ within the spike radius. This, however, is only an optimistic limit and doesn't have to hold. The DM density close to the BH could be practically far lesser, in which case the exclusion limits would be weaker. One can consider the following normalization for DM density profile

\begin{eqnarray}
\int_{4 R_S} ^{R_{\text{sp}}} 4 \pi r^2 \rho (r) dr = x ~M_{\text{BH}}
\end{eqnarray}

{where $x \leq 1$ can be used to reduce number of DM particles in the spike. The DM density profiles and LOS integral for the two profiles $\gamma_{\text{sp}}=7/3,3/2$ and the two BMPs are given in Figs.~\ref{fig:Density_x_LOS_1} and \ref{fig:Density_x_LOS_2}, for three different values of $x$. The resultant exclusion limits are plotted for heavy vector mediator scenario in Fig.~\ref{fig:ExclBound_diff_x}. Obviously, with decreasing number of DM particles in the spike, the exclusion limits become weaker. Clearly, the DM density and hence LOS integral profiles scale as $\sim x$ for BMP1, and hence the exclusion limits for various choices of normalisation scale as $\sim \frac{1}{\sqrt{x}}$ for both profiles. However, for BMP2, the proportionality is not that straightforward. The height ($\rho_{\text{sat}}$) of the DM density plateau at small $r$ is independent of the choice of normalisation, but depends on $m_\chi$. Depending on the mass of DM particle under consideration, there could be a significant number of DM particles left when $x$ is reduced. For lighter DM particles, the DM density fall off behaviour with respect to $x$ is different for Profile 1 and Profile 2, which is just a consequence of the parameterization of the density profile. As a result of this, for lighter DM and BMP2, the qualitative behaviour of the exclusion bounds, as $x$ is reduced by orders of magnitude, is different in the two profiles.}

\begin{figure}
\centering
\begin{tabular}{cc}
\subcaptionbox{DM Density Profiles for BMP1 \label{DMprofilesBMP1}}{\includegraphics[scale=0.6]{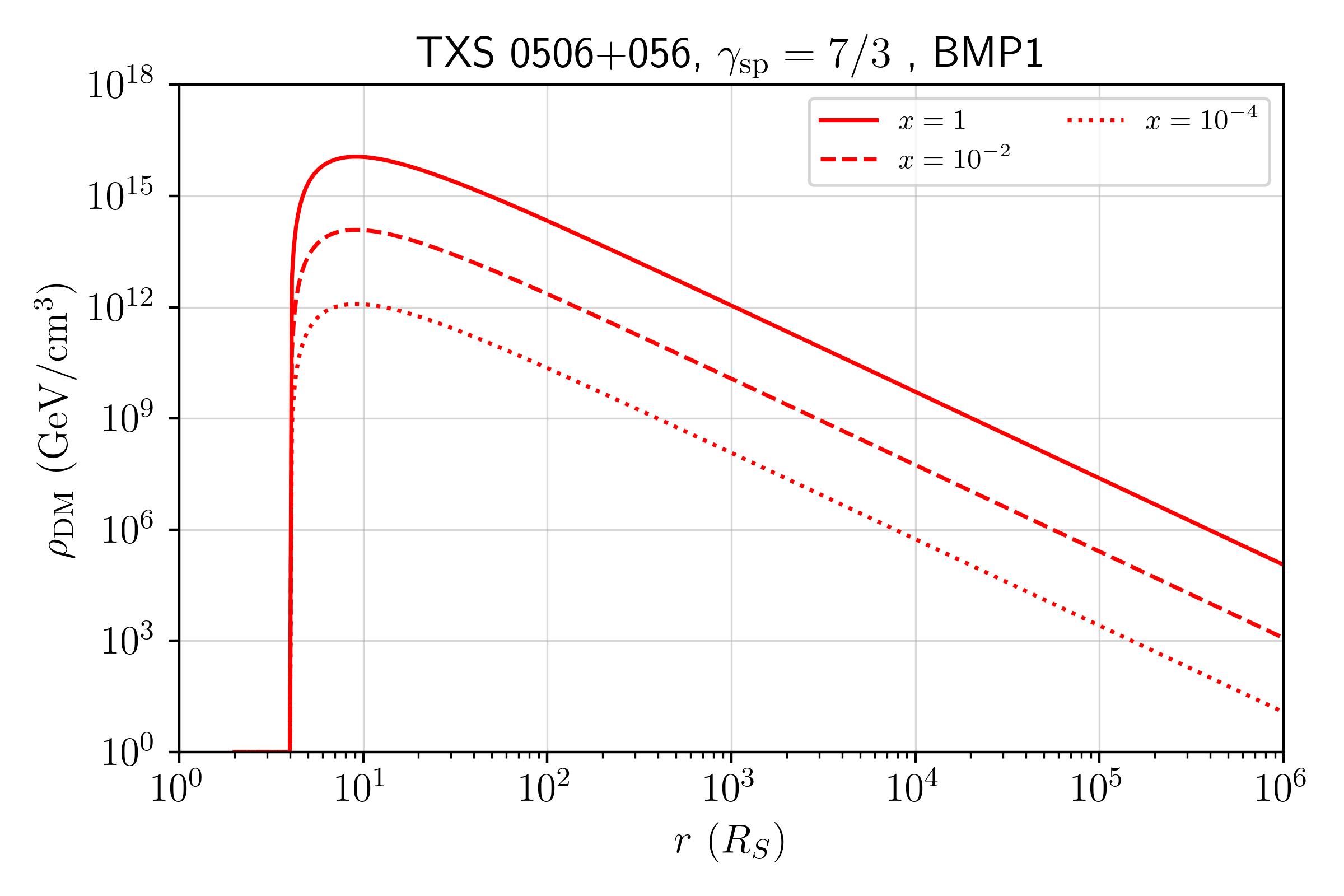}} &
\subcaptionbox{DM Density Profiles for BMP2 \label{DMprofilesBMP2}}{\includegraphics[scale=0.6]{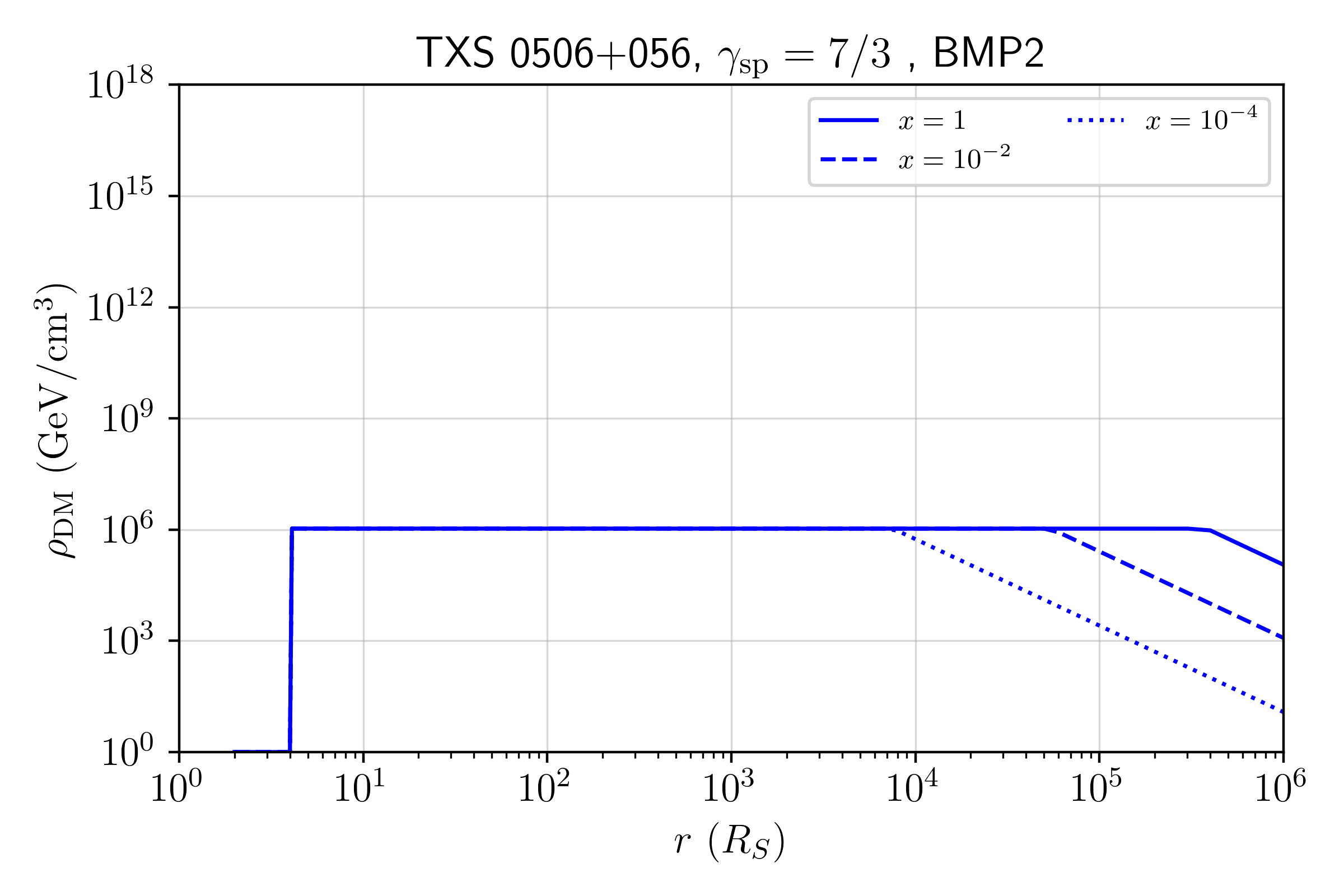}}  \\
\subcaptionbox{LOS integral Profiles for BMP1 \label{LOSprofilesBMP1}}{\includegraphics[scale=0.6]{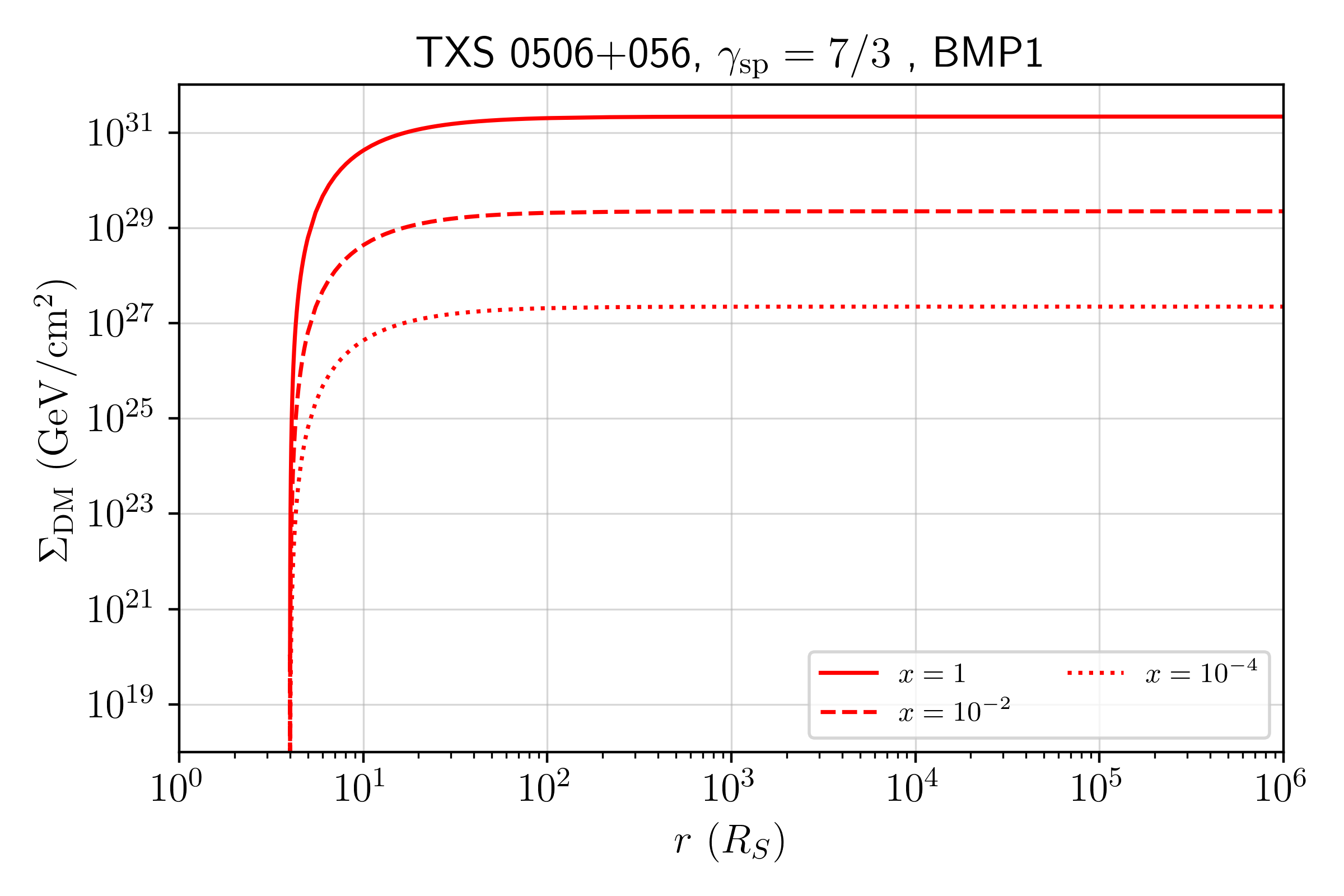}} &
\subcaptionbox{LOS integral Profiles for BMP2 \label{LOSprofilesBMP2}}{\includegraphics[scale=0.6]{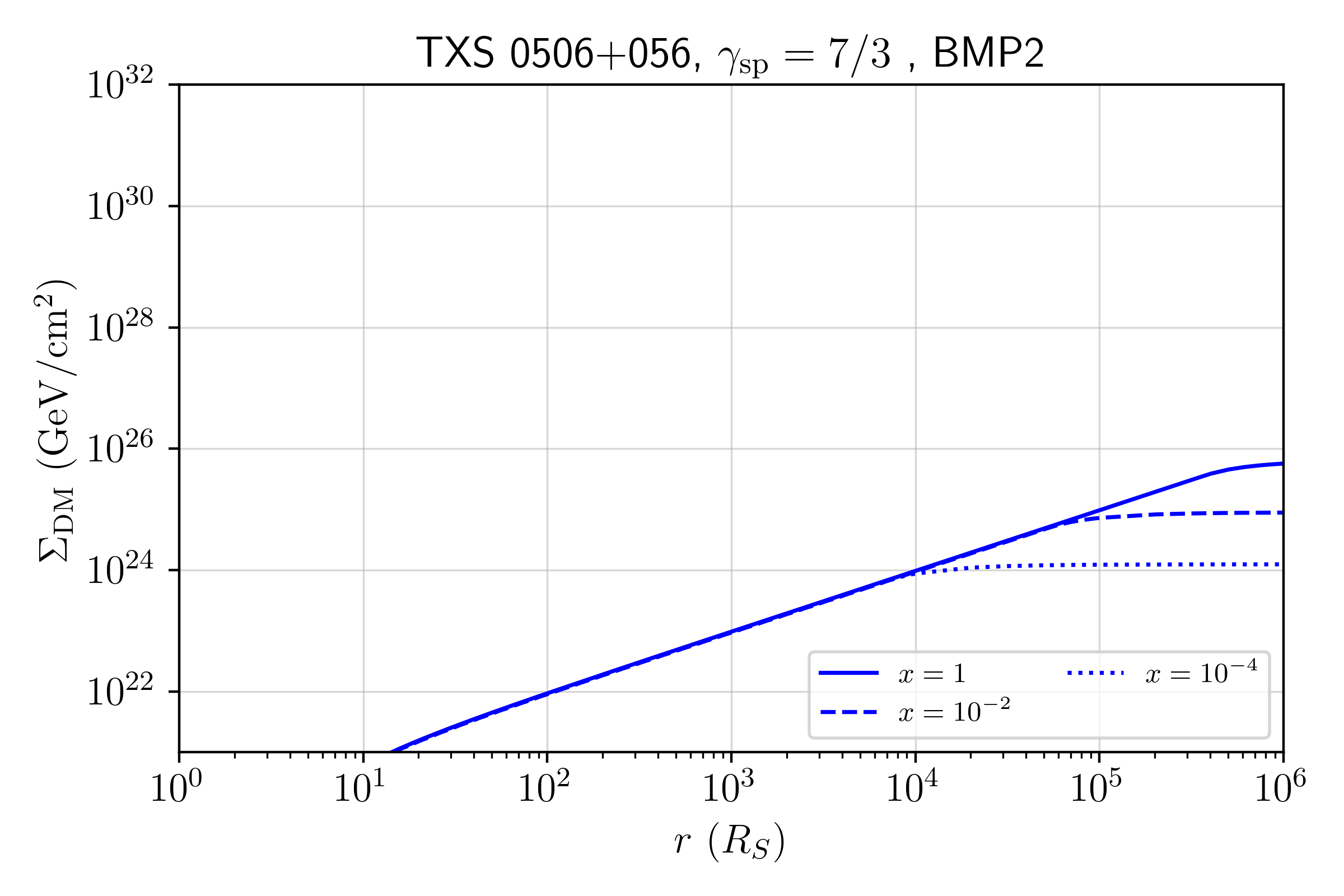}}
\end{tabular}
\caption{The DM density profiles and LOS integral profiles for $\gamma_{\text{sp}}=7/3$ are plotted above for BMP1 (red lines) and BMP2 (blue lines). The values of $x$ chosen are $x=1$ (solid lines), $x=10^{-2}$ (dashed lines) and $x=10^{-4}$ (dotted lines).}
\label{fig:Density_x_LOS_1}
\end{figure}

\begin{figure}
\centering
\begin{tabular}{cc}
\subcaptionbox{DM Density Profiles for BMP1 \label{DMprofilesBMP1}}{\includegraphics[scale=0.6]{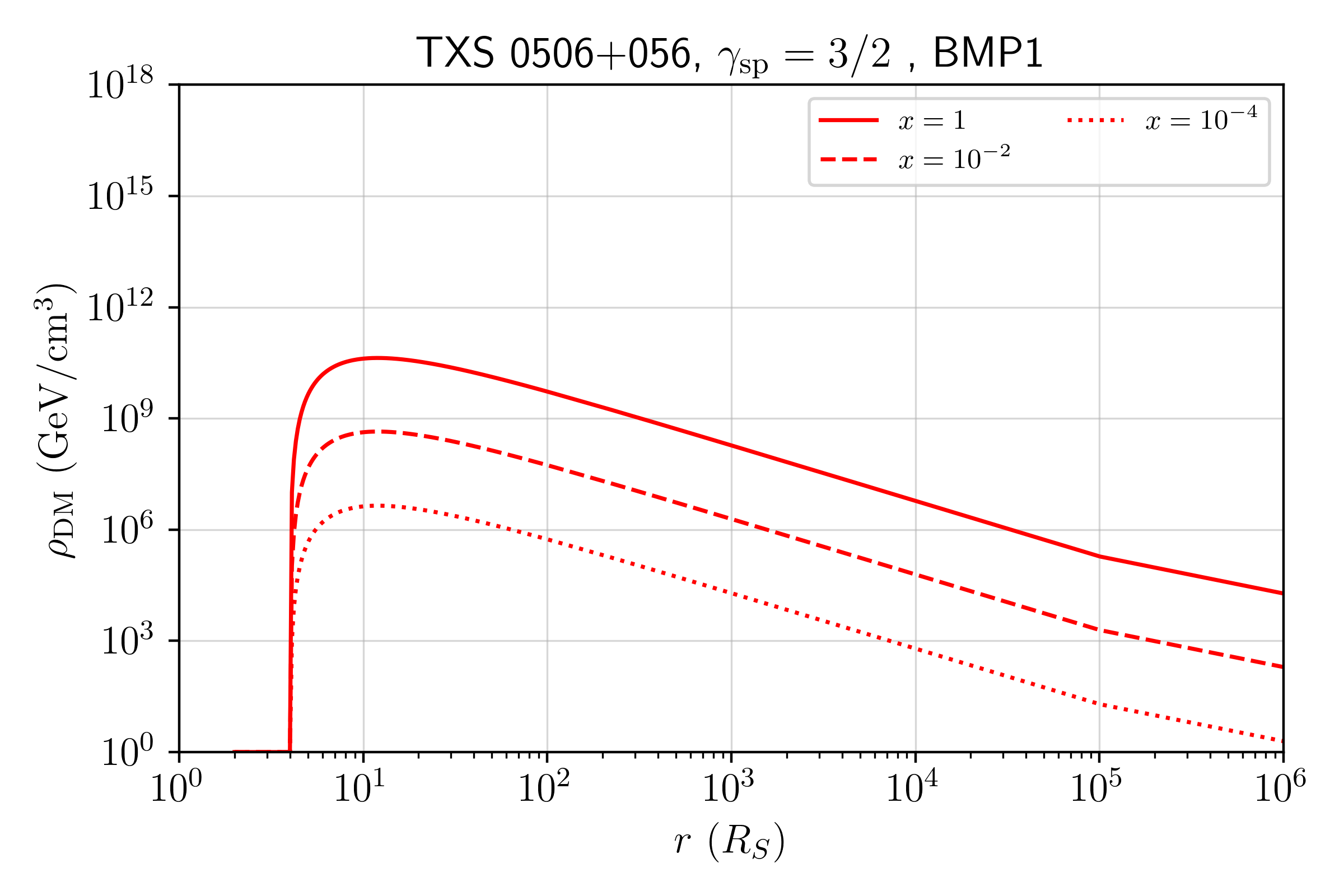}} &
\subcaptionbox{DM Density Profiles for BMP2 \label{DMprofilesBMP2}}{\includegraphics[scale=0.6]{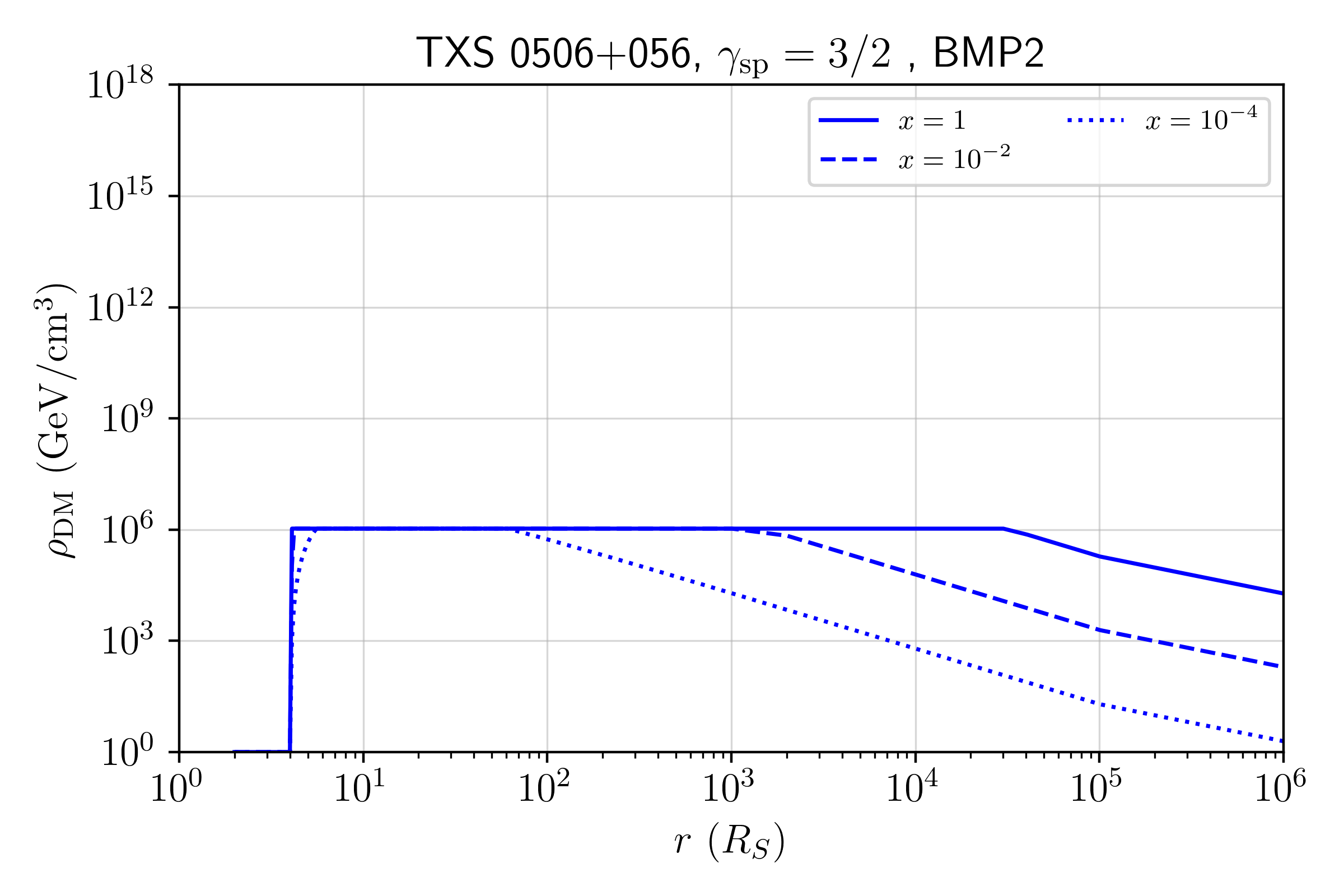}}  \\
\subcaptionbox{LOS integral Profiles for BMP1 \label{LOSprofilesBMP1}}{\includegraphics[scale=0.6]{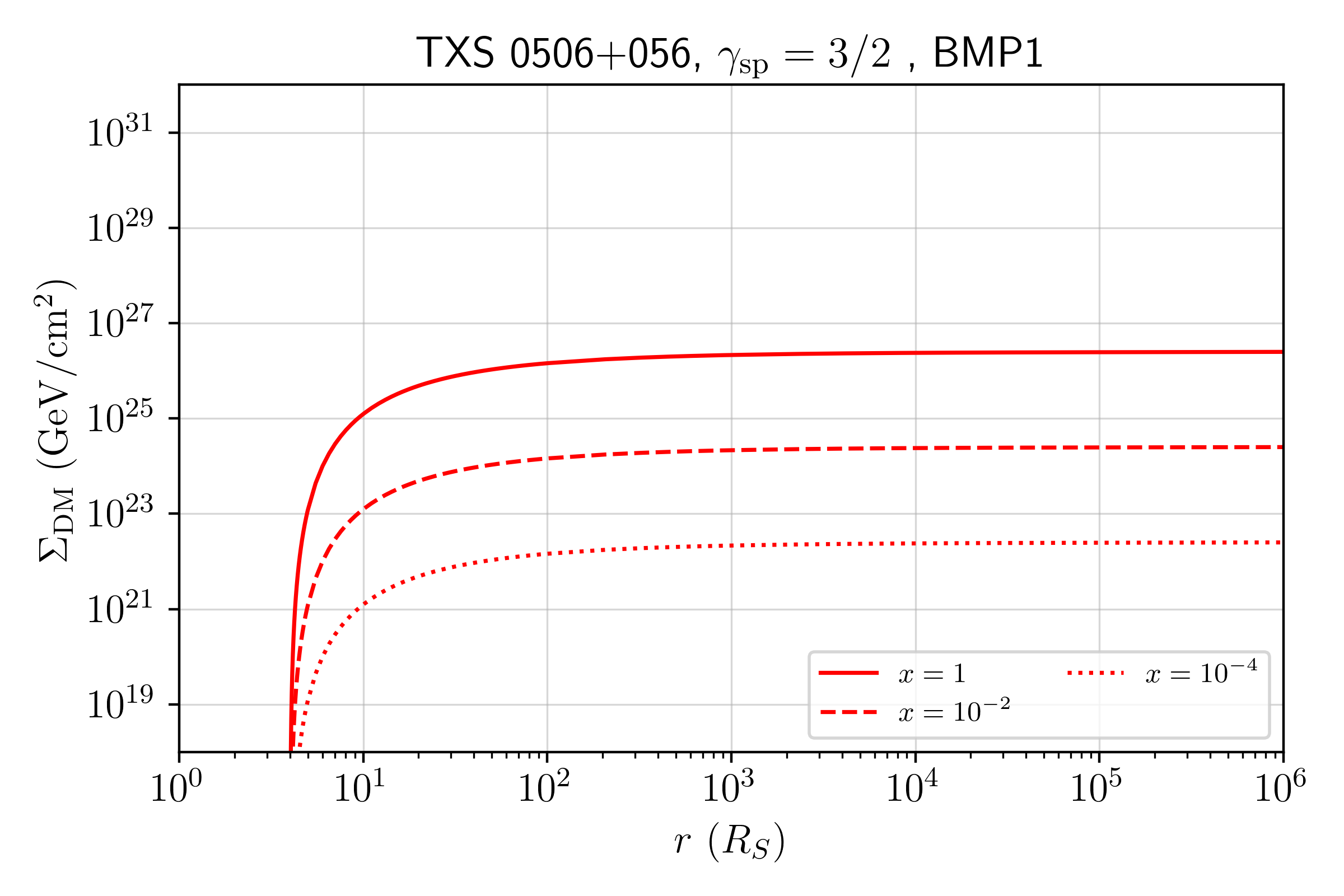}} &
\subcaptionbox{LOS integral Profiles for BMP2 \label{LOSprofilesBMP2}}{\includegraphics[scale=0.6]{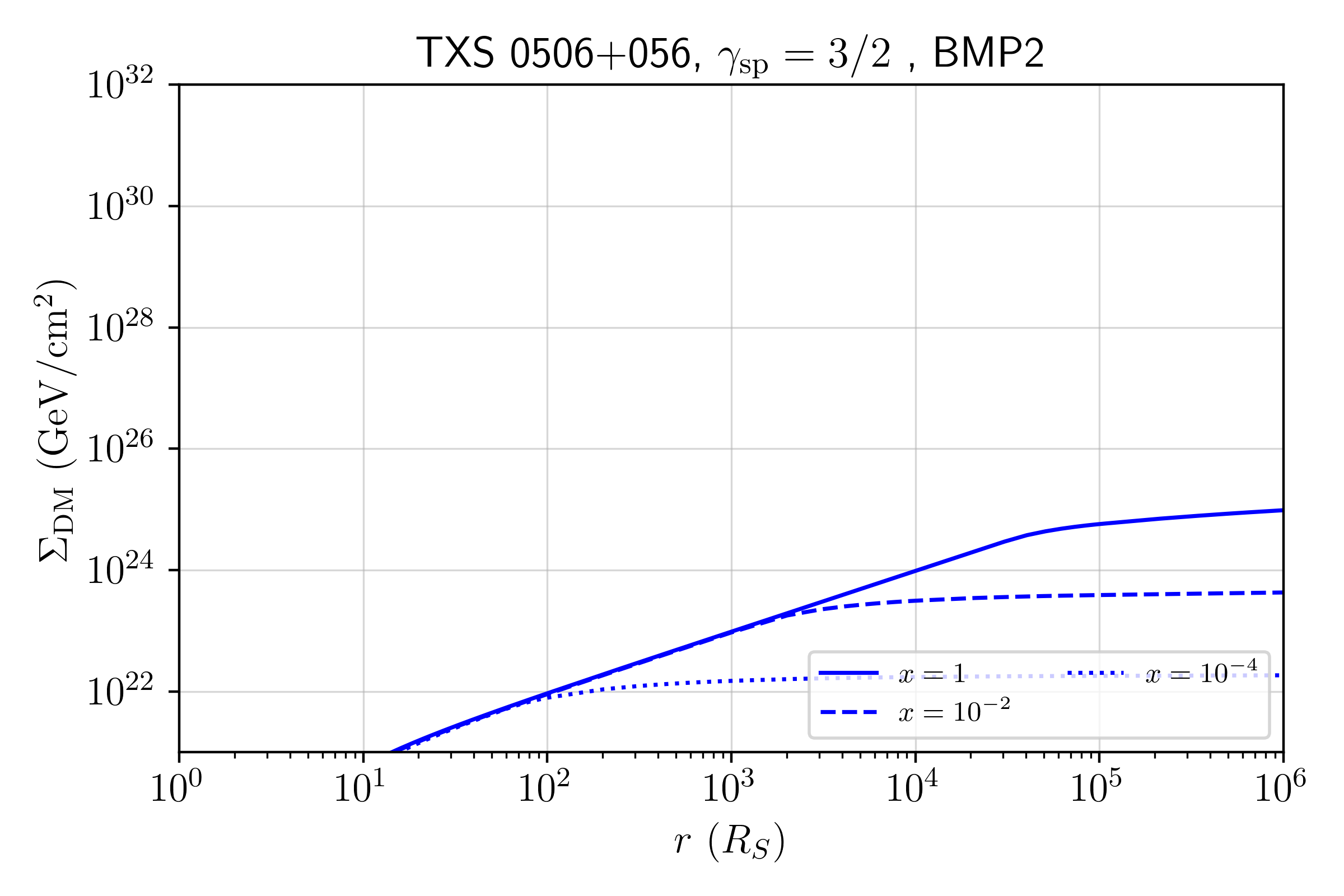}}
\end{tabular}
\caption{The DM density profiles and LOS integral profiles for $\gamma_{\text{sp}}=3/2$ are plotted above for BMP1 (red lines) and BMP2 (blue lines). The values of $x$ chosen are $x=1$ (solid lines), $x=10^{-2}$ (dashed lines) and $x=10^{-4}$ (dotted lines).}
\label{fig:Density_x_LOS_2}
\end{figure}

\begin{figure}[!h]
\centering
\begin{tabular}{cc}
\subcaptionbox{Exclusion bound for Profile 1 \label{fig:ExclBound_diff_x_Prof1}}{\includegraphics[scale=0.6]{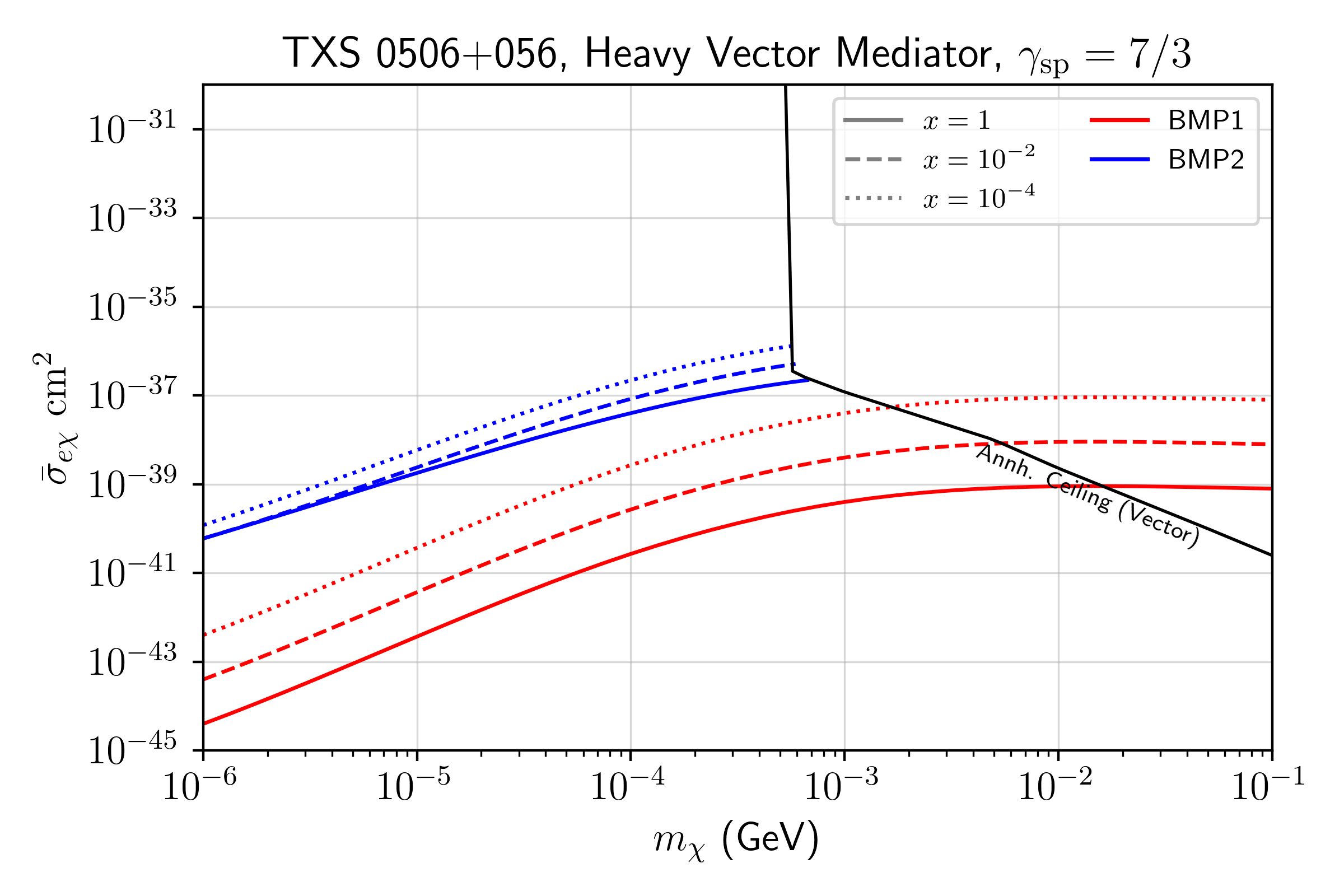}} &
\subcaptionbox{Exclusion bound for Profile 2 \label{fig:ExclBound_diff_x_Prof2}}{\includegraphics[scale=0.6]{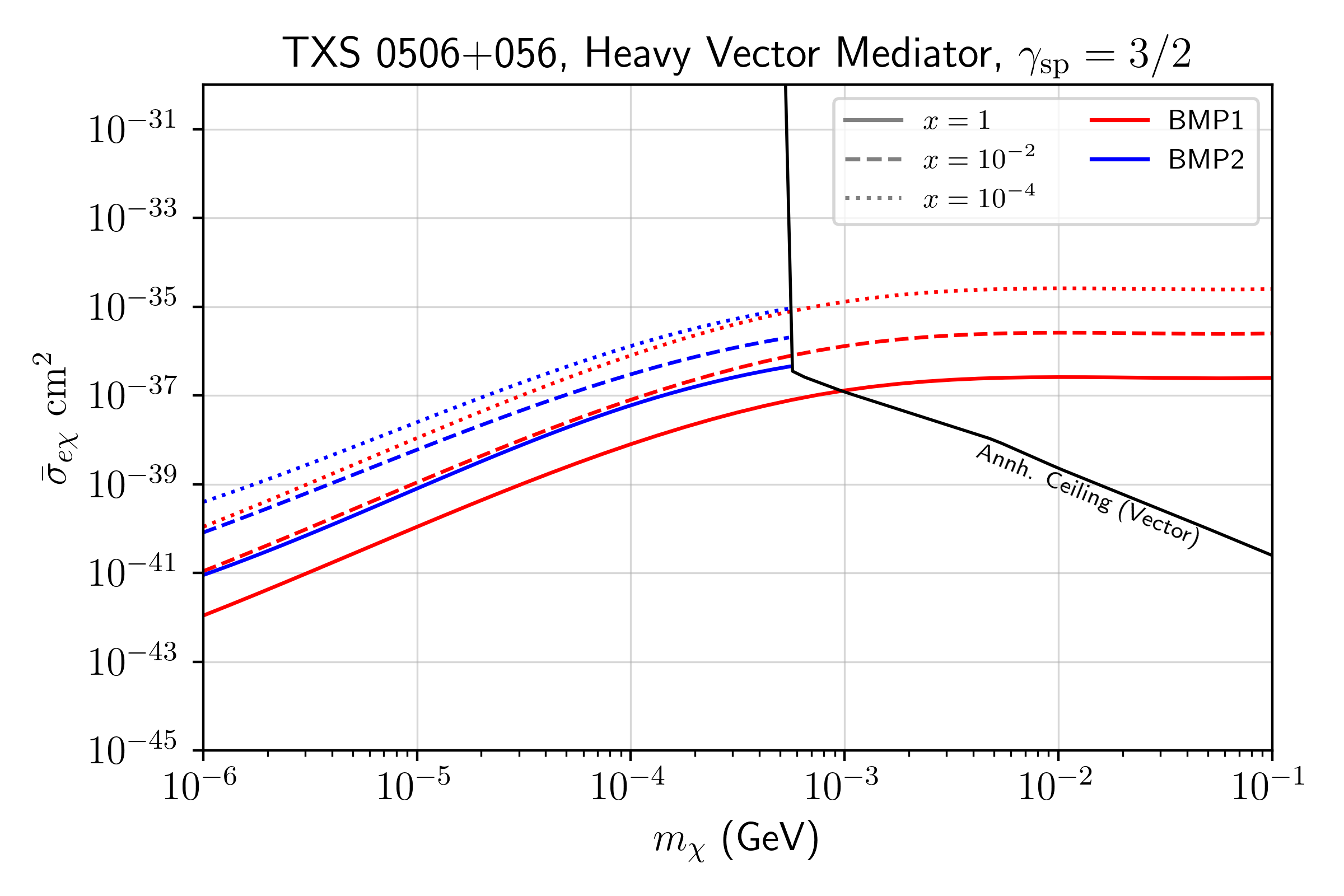}}  \\
\end{tabular}
\caption{The exclusion bound is plotted for the blazar TXS 0506+056 for BMP1 (red lines) and BMP2 (blue lines), corresponding to the heavy vector mediator scenario, for $\gamma_{\text{sp}}=7/3$ (\ref{fig:ExclBound_diff_x_Prof1}) and $\gamma_{\text{sp}}=3/2$ (\ref{fig:ExclBound_diff_x_Prof2}). The values of $x$ chosen are $x=1$ (solid lines), $x=10^{-2}$ (dashed lines) and $x=10^{-4}$ (dotted lines).}
\label{fig:ExclBound_diff_x}
\end{figure} 

\bibliography{reference}
\end{document}